\documentclass[reprint,amsmath,amssymb,aps]{revtex4-1}
\usepackage{ae,aecompl}

\usepackage[LGR,T1]{fontenc}
\usepackage[latin9]{inputenc}
\usepackage{geometry}
\usepackage{float}
\usepackage{amsmath}
\usepackage{amssymb}
\usepackage{graphicx}
\usepackage{esint}
\makeatletter

\usepackage{dcolumn}
\usepackage{bm}
\usepackage{amstext}
\renewcommand\paragraph{\@startsection{paragraph}{4}{\z@}
   {-3.25ex\@plus -1ex \@minus -.2ex}
   {1.5ex \@plus .2ex}%
   {\normalfont\normalsize\bfseries}}

\makeatother

\begin{document}
\preprint{APS/123-QED}\title{Self-avoiding worm-like chain model for dsDNA loop formation}
\author{Yaroslav Pollak}
 \altaffiliation[Also at ]{Russell Berrie Nanotechnology Institute, Technion}
 \author{Sarah Goldberg}
\author{Roee Amit}
\email{roeeamit@technion.ac.il}
 \altaffiliation[Also at ]{Russell Berrie Nanotechnology Institute, Technion}
 \affiliation{Biotechnology and Food Engineering, Technion - Israel Institute of Technology, Haifa, Israel 32000}

\date{\today}

\begin{abstract} We compute for the first time the effects of excluded volume on the probability for double-stranded DNA to form a loop. We utilize a Monte-Carlo algorithm for generation of large ensembles of self-avoiding worm-like chains, which are used to compute the J-factor for varying lengthscales. In the entropic regime, we confirm the scaling-theory prediction of a power-law drop off of $-1.92$, which is significantly stronger than the $-1.5$ power-law predicted
by the non-self-avoiding worm-like chain model. In the elastic regime, we find that the angle-independent end-to-end chain distribution is highly anisotropic. This anisotropy, combined with the excluded volume constraints, lead to an increase in the J-factor of the self-avoiding worm-like chain by about half an order of magnitude relative to its  non-self-avoiding counterpart. This increase could partially explain the anomalous results of recent cyclization experiments, in which short
dsDNA molecules were found to have an increased propensity to form a loop.
\end{abstract}
\maketitle \newcommand{\listequationnumber}{\refstepcounter{equation}(\theequation)}\newcommand{\listequation}[1]{\hfill$\displaystyle #1$\hfill\listequationnumber}

\section{Introduction}

Cyclization or looping of polymers has functioned as an important
experimental paradigm for probing nanometric elastic characteristics
of polymers for more than half a century \cite{jacobson_intramolecular_1950,flory_statistical_1969,crothers_dna_1992,cloutier_dna_2005}.
In particular, experimental studies of polymer cyclization, in conjunction
with modern single-molecule approaches \cite{marko_stretching_1995,wiggins_high_2006},
have been used extensively in recent years to study double-stranded
DNA (dsDNA). While these studies have lead to important advances in
our understanding of the properties of dsDNA polymer physics, a detailed
comparison of the data that has emerged from the different experimental
approaches has generated surprising and unexpected puzzles such as
dsDNA hyper-bendability at short polymer lengths \cite{cloutier_dna_2005,vafabakhsh_extreme_2012,wiggins_high_2006}.
Since many important biological regulatory processes that occur at
the level of the genome are strongly related to DNA looping or cyclization,
it is crucial to develop a detailed theoretical understanding of dsDNA
cyclization that can explain the seemingly contrasting force extension
\cite{marko_stretching_1995} and cyclization \cite{cloutier_dna_2005,vafabakhsh_extreme_2012,wiggins_high_2006}
experimental observations.

As a result of these experimental works, a consensus picture of dsDNA
as a semi-flexible polymer has emerged. At lengths that are smaller
than the Kuhn length \cite{flory_statistical_1969} DNA behaves like
a rigid rod. At intermediate lengths it behaves like a Gaussian chain
\cite{gennes_scaling_1979}. At large lengths, dsDNA becomes a swollen
chain due to excluded volume effects that emerge from the inability
of DNA to cross itself \cite{gennes_scaling_1979,flory_statistical_1969}.
A recent set of theoretical studies \cite{chen_renormalization-group_1992,nepal_structure_2013}
showed that the DNA length, at which the transition from the ideal
to the swollen chain occurs is highly dependent on the effective aspect
ratio of DNA, which we define as the ratio of the effective width
or cross-section of DNA $\mathit{w}$ to the Kuhn length $b$. As
$\frac{w}{b}\rightarrow0$, the DNA transition length becomes larger,
which implies that the polymer behaves like an ideal chain for an
ever increasing length. Alternatively, as $\mathit{\frac{w}{b}\rightarrow\text{1}}$,
the DNA transition length shrinks, and the DNA behaves increasingly
like a swollen chain for all lengths. Since DNA is anisotropic with
an effective aspect ratio that is estimated to be in the range of
$0.03-0.09$~\cite{rybenkov_probability_1993}, the power-law
exponent for the end-to-end distance of DNA was predicted \cite{tree_is_2013}
to not deviate substantially from the Gaussian chain prediction for
lengths as high as $\sim$100 kb. This prediction was recently confirmed
experimentally, showing a radius of gyration power-law of $0.52\pm0.02$
\cite{nepal_structure_2013}, that is very close to the Gaussian chain
prediction of 0.5.

The process of DNA cyclization or looping has not been modeled to
date by algorithms that take into consideration the effective aspect
ratio of the DNA. Historically, this process is quantified by the
Jacobson-Stockmayer factor \cite{jacobson_intramolecular_1950}, which
was defined as \cite{crothers_dna_1992}

\begin{equation}
J=\frac{K_{C}}{K_{D}},
\end{equation}
namely, the ratio of the equilibrium constant of cyclization $K_{C}$
to that of bimolecular association $K_{D}$, assuming the measurement
is carried out on a dilute solution of dsDNA in standard saline conditions
\cite{crothers_dna_1992}. This definition led Flory, Shimada, and
Yamakawa \cite{flory_macrocyclization_1966,yamakawa_modern_1971}
to re-derive the J-factor via equilibrium thermodynamic considerations
as the probability for the two ends of the dsDNA to bond, normalized
by the infinitesimal bonding volume and angular tolerance in order
to avoid any measurable ambiguities that might depend on arbitrary
choices of the bonding volume.

Experimentally, measuring the J-factor for DNA has proven to be quite
challenging, but improved technology has yielded a set of increasingly
more precise \textit{in vivo} \cite{amit_building_2011} and \textit{in
vitro} measurements \cite{cloutier_dna_2005,vafabakhsh_extreme_2012}
of this quantity. As a result, an interesting if not controversial
picture had emerged. While the behavior for longer lengths $\mathit{(L>b)}$
has matched well with the predictions of the worm-like chain (WLC)
model, for shorter chains where $\mathit{L<b}$ a deviation from the
WLC model has been consistently observed \cite{cloutier_dna_2005,wiggins_high_2006,vafabakhsh_extreme_2012,amit_building_2011}.
In this regime, the dsDNA is expected to behave like a rigid rod whose
propensity to form a loop depends strongly on the elastic or bending
energy. Therefore, as the chain becomes shorter, the probability to
form a loop is expected to decrease exponentially. However, the experimental
data showed that this prediction is not observed to the extent predicted
by the WLC model, and instead an increasing propensity for bending
as compared with model predictions was detected as the length of the
DNA chain was decreased below the Kuhn length. Several theoretical
hypotheses were raised to explain this discrepancy, including localized
DNA kinking or formation of melting bubbles \cite{yan_statistics_2005,wiggins_exact_2005},
and possible sub-harmonic contributions to the WLC which render the
DNA more flexible at short lengths \cite{wiggins_generalized_2006}.
However, these modifications have not been proven conclusively to
be the underlying cause for the experimental results, and the problem
remains open.

In this work we study cyclization of DNA with finite width. Section~\ref{sec:Theory} introduces the basic theory that underlies the definition of the
self-avoiding J-factor integral. We utilize an advanced Monte Carlo
algorithm, described in Sec.~\ref{sec:numerical-implementation}, that is based on the weighted
Rosenbluth and Rosenbluth method, similar to the algorithm used by
Tree et al.,\cite{tree_is_2013} to generate ensembles of anisotropic
DNA chains up to $5000$~bp in length. Our algorithm is able to reproduce
the theoretically-predicted behavior of DNA at long lengths. In Sec.~\ref{sec:results} we use our algorithm to compute for the first time the J-factor
for self-avoiding polymers. In particular, for long chain lengths
the numerically computed J-factor power-law exponent converges on
the scaling law prediction~\cite{gennes_scaling_1979,sinclair_cyclization_1985}
of $-1.92$ regardless of the width of the chain. For shorter lengths
we find that the chosen definition of the end-to-end bonding strongly
affects the J-factor behavior. We demonstrate that this function
can be systematically changed by choosing various end-to-end bonding
conditions, and lead to an apparent increased ``bendability'' effect
in this regime as compared with the WLC model.

\section{\label{sec:Theory}Theory}

\subsection{\label{sub:The-anisotropic-WLC}The self-avoiding worm-like chain
(SAWLC) model}

DNA is typically modeled as a discrete semi-flexible chain made of
individual and irreducible links of length $l$, such that the deviation
of one link from its neighboring link depends strongly on some elastic
energy. This class of polymer models is based on the original work
of Kratky and Porod \cite{o._kratky_g._porod_rontgenuntersuchung_1949}
and is referred to as the class of worm-like chain (WLC) models. However,
except for a few notable exceptions \cite{chen_renormalization-group_1992,tree_is_2013},
the WLC models do not take into account energetic and entropic effects
that emerge from the cross-section or ``thickness'' of the DNA double
helix.

\begin{figure}
\includegraphics[width=1\columnwidth]{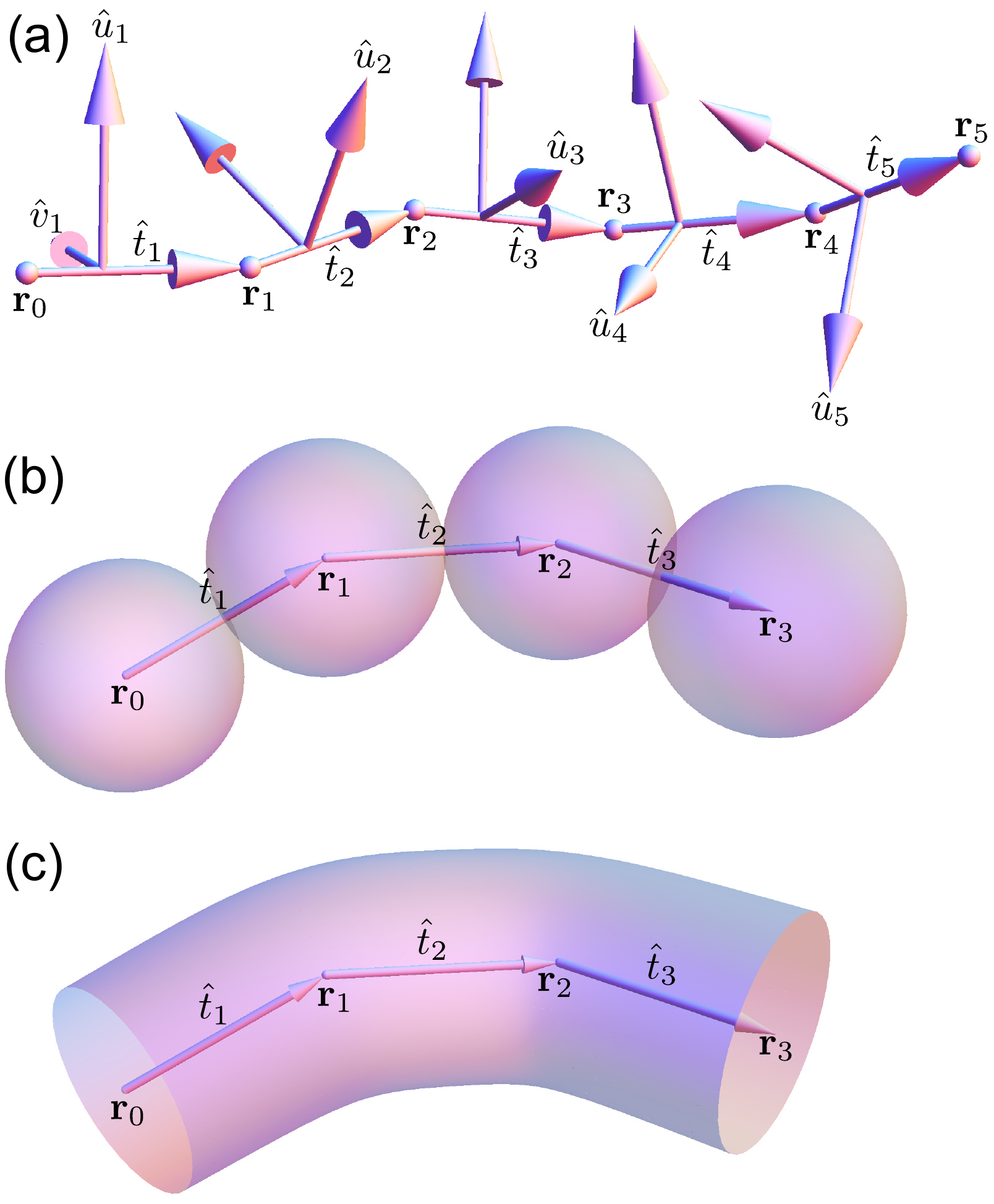}
\caption{\label{fig:WLC-SAWLC} (Color online) WLC and SAWLC chains.
A discrete chain is described by the locations of its joints $\mathbf{r}_{i}$,
and a local coordinate system defined by three orthonormal vectors
$\hat{u_{i}},\hat{v}_{i},\hat{t}_{i}$ at each link. The vector $\hat{t}_{i}$
points along the direction of the ith link. (a) A WLC chain with 5
links and link length $l=1$. (b) A SAWLC chain with 3 links and $l=w=1$.
$\hat{u}_{i}$ and $\hat{v}_{i}$ are not shown. (c) Representation
of an ideally modeled chain with finite width.}
\end{figure}

In the following, we describe each chain by the locations of its elements,
and a local coordinate system defined by three orthonormal vectors
$\hat{u},\hat{v},\hat{t}$ at each element. The vector $\hat{t}$
points along the direction of the chain. For the continuous
WLC these vectors are defined continuously along the chain contour.
For the discrete WLC, the joint locations $\mathbf{r}_{i}$ and the
local coordinate systems of the links fully define the chain. We number
the links of a chain in the range $1..N$ for a chain of $N$ links,
and the end-points of the links (chain joints) in the range $0..N$,
where joint 0 is the beginning terminus of the chain (see Fig.~\ref{fig:WLC-SAWLC}(a)).
When mapping worm-like chains to actual dsDNA, the chain links correspond
to DNA base-pairs and chain joints correspond to mid-points between
DNA base-pairs.

The conventional bending energy for the WLC models is the elastic
energy associated with bending link $i\in\left\{ 2,\,...,\, N\right\} $ relative to link $i-1$
with angles $\theta_{i},\phi_{i}$ (zenith and azimuthal angles in
local spherical coordinates of link $i-1$),
which can be written as
\begin{equation}
\beta E^{el}\left(\theta_{i},\phi_{i}\right)=\frac{a}{2}\left|\hat{t}_{i}-\hat{t}_{i-1}\right|^{2}=a\left(1-\cos\theta_{i}\right),\label{eq:WLC_Bending_E}
\end{equation}
where $a$ corresponds to the bending rigidity of the DNA chain (assuming
azimuthal symmetry), and $\beta=\left(k_{B}T\right)^{-1}$. It is important to note that the angles $\theta_{i},\phi_{i}$ are
given in the \textit{local coordinate system} of the $\left(i-1\right)$th
link. For a specific configuration of the chain, we
introduce the notation $\{\theta_{n},\phi_{n}\}$ to denote the set
of all the links' angles of the chain, from link 1 to link $n$.

To account for the finite thickness of the DNA we introduce a second
energy contribution. We engulf each joint by a ``hard-wall'' spherical
shell of diameter $w$. This allows us to model the final contribution
to the elastic energy as a set of hard-wall potentials. For the simple
case in which the chain link length is larger than the link diameter
($l\geq w$), and therefore no two neighboring hard-wall spheres overlap,
the hard-wall potential energy for the $i$th chain link can be defined
as
\begin{equation}
E_{i}^{hw}\left(\left\{ \theta_{i},\phi_{i}\right\} \right)=\begin{cases}
\infty & \,\,\,\,\mbox{joint \ensuremath{i} overlaps with one}\\
 & \,\,\,\,\mbox{or more joints }0..\left(i-1\right)\\
0 & \,\,\,\,\mbox{otherwise}
\end{cases}.\label{eq:E_i^hw}
\end{equation}
This allows us to write an expression for the total elastic energy
associated with the chain of spheres as follows:
\begin{equation}
E\left(\left\{ \theta_{N},\phi_{N}\right\} \right)=\sum_{i=1}^{N}E^{el}\left(\theta_{i},\phi_{i}\right)+\sum_{i=1}^{N}E_{i}^{hw}\left(\left\{ \theta_{i},\phi_{i}\right\} \right).\label{eq:SAWLC_E}
\end{equation}
We emphasize that this approach (Fig.~\ref{fig:WLC-SAWLC}(b)) is an
approximation for the more realistic uniformly thick chain (Fig.~\ref{fig:WLC-SAWLC}(c)).
We choose it for our model due to its simplicity and adequate approximation
of the chain excluded volume. However, it may result in incorrect
representation of chain excluded volume near the chain termini. In
particular, if we compare Fig.~\ref{fig:WLC-SAWLC}(c) to Fig.~\ref{fig:WLC-SAWLC}(b),
we note that the minimally possible end-to-end distance of the chain
is $0$ in the former and $w$ in the latter representations, respectively.

The configurational partition function for the model DNA chain consisting
of $N$ links is defined as
\begin{eqnarray}
Z_{N} & = & \int\limits _{-1}^{1}\mbox{d}\cos\theta_{1}\int\limits _{0}^{2\pi}\mbox{d}\phi_{1}\cdots\nonumber \\
 &  & \!\!\!\!\!\int\limits _{-1}^{1}\mbox{d}\cos\theta_{N}\int\limits _{0}^{2\pi}\mbox{d}\phi_{N}\exp\left[-\beta E\left(\left\{ \theta_{N},\phi_{N}\right\} \right)\right],\label{eq:SUP-General_Z}
\end{eqnarray}
where $\beta=\frac{1}{k_{b}T}$. Substituting $E_{i}^{hw}$ from Eq.~(\ref{eq:E_i^hw})
and opening the sums yields:
\begin{widetext}
\begin{equation}
Z_{N} =  \!\!\int\limits _{-1}^{1}\!\!\mbox{d}\cos\theta_{1}\!\int\limits _{0}^{2\pi}\!\!\mbox{d}\phi_{1}\exp\left[-\beta E_{1}^{el}\left(\theta_{1},\phi_{1}\right)\right]\Theta_{1}^{hw}\left(\left\{ \theta_{1},\phi_{1}\right\} \right)\cdots \!\! \int\limits_{-1}^{1}\!\!\mbox{d}\cos\theta_{N}\!\int\limits _{0}^{2\pi}\!\!\mbox{d}\phi_{N}\exp\left[-\beta E_{1}^{el}\left(\theta_{N},\phi_{N}\right)\right]\Theta_{N}^{hw}\left(\left\{ \theta_{N},\phi_{N}\right\} \right),\label{eq:SUP-MC_Z-1}
\end{equation}
\end{widetext}
where
\begin{equation}
\Theta_{i}^{hw}\left(\left\{ \theta_{i},\phi_{i}\right\} \right)=\begin{cases}
0 & \,\,\,\,\mbox{joint \ensuremath{i} overlaps with one}\\
 & \,\,\,\,\mbox{or more joints}\,0..\left(i-1\right)\\
1 & \,\,\,\,\mbox{otherwise}
\end{cases}.\label{eq:Theta-HW}
\end{equation}

We term the model described by Eq.~(\ref{eq:SAWLC_E}) the self-avoiding
worm-like chain model (SAWLC).

\subsection{\label{sub:Deriving-the-J-factor}Generalizing the J-factor for the SAWLC}

The process of cyclization \cite{jacobson_intramolecular_1950}, defined
as the joining of one end of the molecule upon itself, can be quantified
by the equilibrium constant for the following process \cite{jacobson_intramolecular_1950,flory_macrocyclization_1966,flory_macrocyclization_1976}:
\begin{equation}
M_{2N}\rightleftharpoons M_{N}+cM_{N},\label{eq:Cyclization_process}
\end{equation}
where $\mathit{M_{N},}\, M_{2N}$ and $\mathit{cM_{N}}$ correspond
to the monomeric, dimeric (defined as the end-to-end joining of two
separate molecules), and cyclized polymer with $N$ links, respectively.
This process can be equivalently expressed as two intermediate processes:\begin{subequations}
\begin{eqnarray}
M_{N}+M_{N} & \rightleftharpoons & M_{2N}\,,\label{eq:Cyclization_subprocess1}\\
M_{N} & \rightleftharpoons & cM_{N}\,.\label{eq:Cyclization_subprocess2}
\end{eqnarray}
\end{subequations}The Jacobson-Stockmayer factor (or J-factor) is
defined as the equilibrium constant for the entire process (\ref{eq:Cyclization_process})
\cite{jacobson_intramolecular_1950,flory_macrocyclization_1976}:
\begin{equation}
J\equiv\frac{\left[M_{N}\right]\left[cM_{N}\right]}{\left[M_{2N}\right]},
\end{equation}
and the equilibrium constants for the intermediate processes are similarly
defined as
\begin{eqnarray}
K_{D} & \equiv & \frac{\left[M_{2N}\right]}{\left[M_{N}\right]^{2}},\label{eq:dimerization Kd}
\end{eqnarray}
and
\begin{equation}
K_{C}\equiv\frac{\left[cM_{N}\right]}{\left[M_{N}\right]},
\end{equation}
which implies that
\begin{equation}
J\equiv\frac{K_{C}}{K_{D}}.\label{eq:J factor}
\end{equation}
Eq.~\ref{eq:J factor} is a useful expression for the J-factor from an
experimental perspective, but less convenient for numerical simulation.

\begin{figure}
\includegraphics[width=1\columnwidth]{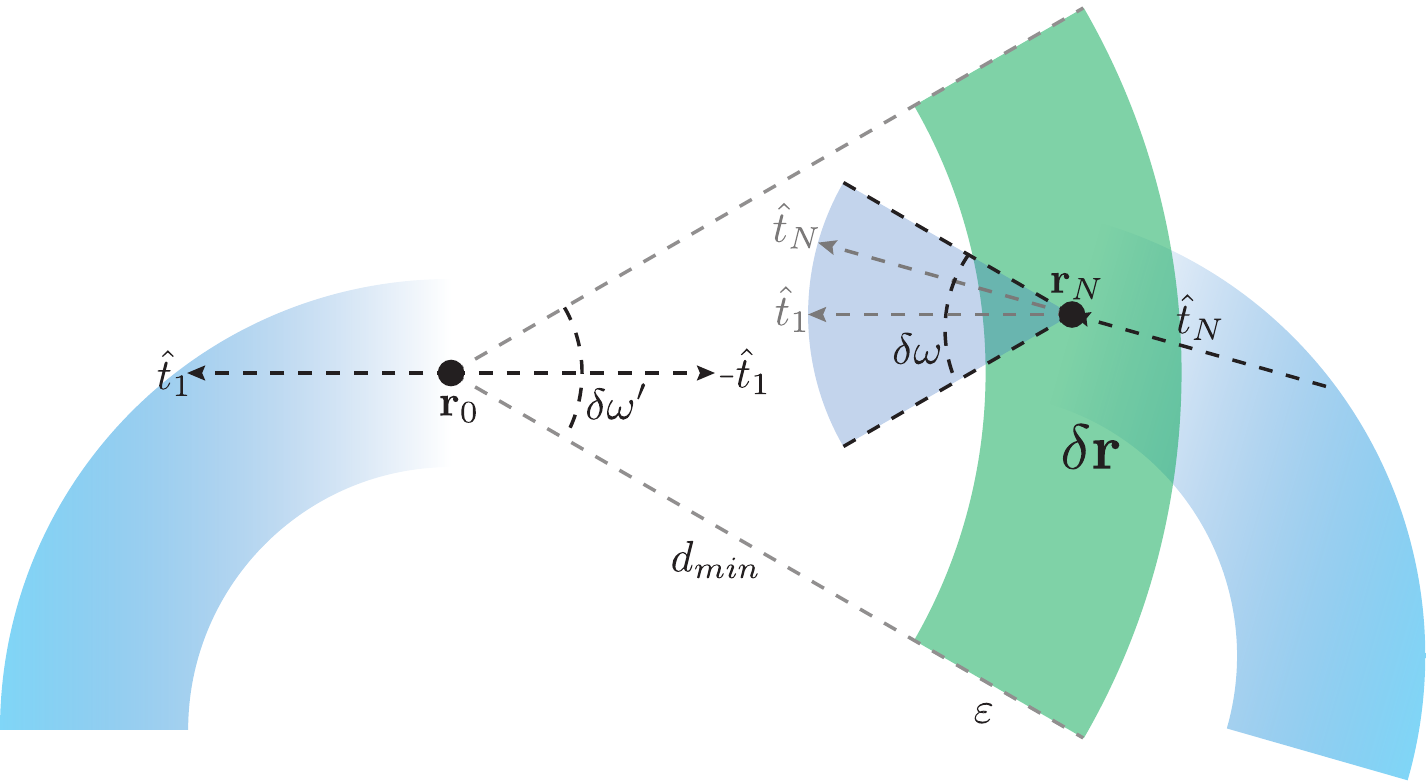}
\caption{\label{fig:bond-formation-criteria}(Color online) Illustration of bond formation
criteria. The termini and their bond directions are denoted
by $\mathbf{r}_{0},\mathbf{r}_{N}$ and $\hat{t}_{1},\hat{t}_{N}$
respectively. The minimal possible separation between the termini
is denoted by $d_{min}$. The cross-section of the volume $\delta\mathbf{r}$
in which $\mathbf{r}_{N}$ must reside is shown as a green (dark gray) wedge.
The possible directions for $\hat{t}_{N}$ are shown as a blue (light gray)
cone with its base at $\mathbf{r}_{N}$. The rest of the chain is
shown schematically with blue (light gray) arcs.}
\end{figure}

In order to derive a more insightful expression for $J$, we must
first set a geometrical definition for bond formation between the
two termini in a chain of finite thickness. We denote the locations
of the bonding termini by $\mathbf{r}_{0}$ and $\mathbf{r}_{N}$
, and the directions of their bonds as $\hat{t}_{1}$ and $\hat{t}_{N}$
(Fig.~\ref{fig:WLC-SAWLC}), regardless of whether these termini
are part of the same chain (i.e. cyclization) or different chains
(i.e. dimerization). We define the minimal possible separation between
the two termini as $d_{min}$. This leads to the following general
bond formation conditions for the SAWLC model (Fig.~\ref{fig:bond-formation-criteria}):
\begin{itemize}
\item $\mathbf{r}_{N}$ is confined to a volume $\delta\mathbf{r}$ around
$\mathbf{r}_{0}$, which is defined by a thin shell between $d_{min}$
and $d_{min}+\varepsilon$ and a solid angle $\delta\omega'$ around
$-\hat{t}_{1}$.\listequation{}\label{enu:bond-formation-1}
\item $\hat{t}_{N}$ is collinear with $\hat{t}_{1}$ within the range $\delta\omega$.\listequation{}\label{enu:bond-formation-2}
\end{itemize}

This definition is an extension of the corresponding bond-formation
condition defined by Flory \cite{flory_statistical_1969} for the
WLC. Note that our definition reduces to the Flory criterion when
we set $d_{min}=0\,\mbox{and}\,\delta\omega'=4\pi$. The motivation
for extending the Flory condition arises from our imperfect understanding
of DNA bond formation. For example, if the transient bonds formed
in (\ref{eq:Cyclization_subprocess1}) and (\ref{eq:Cyclization_subprocess2})
are identical to the bonds that comprise the bulk of the chain, then
the correct representation of the chain would have to be the one presented
in Fig.~\ref{fig:WLC-SAWLC}(c) with $d_{min}=0$. If, however, the
termini have a repulsive interaction below some separation $d_{min}$,
then $d_{min}>0$, as shown for the hard-wall interaction
in Fig.~\ref{fig:WLC-SAWLC}(b).

Using our geometrical definition for bonding, we can now consider
the process in (\ref{eq:Cyclization_subprocess1}). Before dissociation
occurs, $\mathbf{r}_{N}$ is confined to a volume $\delta\mathbf{r}$
around $\mathbf{r}_{0}$. In that volume the orientation of the bond
direction $\hat{t}_{N}$ is further restricted by a solid angle of
$\delta\omega$ around $\hat{t}_{1}$. After the dissociation, it
can be anywhere in volume $v$, and its bond can assume any angle
within $4\pi$. Denoting the symmetry number of a cyclic chain species
$\sigma_{a}$, we obtain the following change in configurational entropy
due to process (\ref{eq:Cyclization_subprocess1}):
\begin{equation}
\Delta S_{\left(a\right)}=R\ln\left(\frac{4\pi v/\sigma_{a}}{\delta\mathbf{r}\delta\omega}\right)=R\ln\left(\frac{V}{N_{A}\sigma_{a}\delta\mathbf{r}\frac{\delta\omega}{4\pi}}\right),
\end{equation}
where we denote the change in entropy as $\Delta S_{\left(a\right)}$,
and the volume per mole as $V$ resulting in $v=\frac{V}{N_{A}}$
as the volume per one molecule, where $N_{A}$ is Avogadro's number. If, for simplicity, we assume that
the species are present in standard state concentration of one mole
per unit volume ($V=1$), we obtain
\begin{equation}
\Delta S_{\left(a\right)}=R\ln\left(\frac{1}{N_{A}\sigma_{a}\delta\mathbf{r}\frac{\delta\omega}{4\pi}}\right).
\end{equation}
The standard molar Gibbs free-energy change for the dimerization process
(\ref{eq:Cyclization_subprocess1}) can be written as
\begin{equation}
\Delta G_{\left(a\right)}^{\circ}=\Delta H_{\left(a\right)}^{\circ}-RT\ln\left(\frac{1}{N_{A}\sigma_{a}\delta\mathbf{r}\frac{\delta\omega}{4\pi}}\right),
\end{equation}
where $\Delta H_{\left(a\right)}^{\circ}$ is the dissociation heat
of the dimer.

In the cyclization process (\ref{eq:Cyclization_subprocess2}) an
intramolecular bond is formed which is equivalent to the one severed
in the dimerization (\ref{eq:Cyclization_subprocess1}). We assume
here that $\Delta H_{\left(b\right)}^{\circ}=-\Delta H_{\left(a\right)}^{\circ}$
where $\Delta H_{\left(b\right)}^{\circ}$ is the dissociation heat
of the ring. This assumption is valid unless the overall length of
the chain is so small as to induce strain\cite{flory_statistical_1969}. For the computation of
the change in configurational entropy due to process (\ref{eq:Cyclization_subprocess2}),
$\Delta S_{\left(b\right)}$, we note that the termini of chain $M_{N}$
must meet within the same ranges of $\delta\mathbf{r}$ and $\delta\omega$
as defined in Eqs.~(\ref{enu:bond-formation-1})-(\ref{enu:bond-formation-2}).
This implies that the probability for the termini to meet is given
by
\begin{equation}
\int\limits _{\delta\mathbf{r}}\int\limits _{\delta\omega}\mathbf{C}\left(\mathbf{r},\omega\right)\mbox{d}\mathbf{r}\mbox{d}\omega,
\end{equation}
where $\mathbf{C}\left(\mathbf{r},\omega\right)$ is the function
expressing the distribution of the end-to-end vector $\mathbf{r}\equiv\mathbf{r}_{N}-\mathbf{r}_{0}$
within a solid angle $\omega$ around $\hat{t}_{1}$ ($\frac{\omega}{2\pi}\equiv\hat{t}_{1}\cdot\hat{t}_{N}$),
per unit range in $\mathbf{r}$ $\left(\int\limits _{v}\int\limits _{4\pi}\mathbf{C}\left(\mathbf{r},\omega\right)d\mathbf{r}d\omega=1\right)$.
Thus, the change in configurational entropy in process (\ref{eq:Cyclization_subprocess2})
is
\begin{equation}
\Delta S_{\left(b\right)}=R\ln\left(\int\limits _{\delta\mathbf{r}}\int\limits _{\delta\omega}\mathbf{C}\left(\mathbf{r},\omega\right)\mbox{d}\mathbf{r}\mbox{d}\omega\frac{\sigma_{a}}{\sigma_{R_{N}}}\right),
\end{equation}
where $\sigma_{R_{N}}$ is the symmetry number of a ring of $N$ links.
Hence
\begin{equation}
\Delta G_{\left(b\right)}^{\circ}=\Delta H_{\left(b\right)}^{\circ}-RT\ln\left(\int\limits _{\delta\mathbf{r}}\int\limits _{\delta\omega}\mathbf{C}\left(\mathbf{r},\omega\right)\mbox{d}\mathbf{r}\mbox{d}\omega\,\frac{\sigma_{a}}{\sigma_{R_{N}}}\right).
\end{equation}
Adding the Gibbs free energy change for processes (\ref{eq:Cyclization_subprocess1})
and (\ref{eq:Cyclization_subprocess2}), we obtain the change in the
Gibbs free energy for the entire process (\ref{eq:Cyclization_process}):
\begin{eqnarray}
\Delta G & = & -T\left(\Delta S_{\left(a\right)}+\Delta S_{\left(b\right)}\right)\nonumber \\
 & = & -RT\ln\left(\frac{\int\limits _{\delta\mathbf{r}}\int\limits _{\delta\omega}\mathbf{C}\left(\mathbf{r},\omega\right)\mbox{d}\mathbf{r}\mbox{d}\omega}{N_{A}\sigma_{R_{N}}\delta\mathbf{r}\frac{\delta\omega}{4\pi}}\right).
\end{eqnarray}
This allows us to extract the J-factor:
\begin{equation}
J=\frac{4\pi\int\limits _{\delta\mathbf{r}}\int\limits _{\delta\omega}\mathbf{C}\left(\mathbf{r},\omega\right)\mbox{d}\mathbf{r}\mbox{d}\omega}{N_{A}\sigma_{R_{N}}\delta\mathbf{r}\delta\omega}.\label{eq:J_factor}
\end{equation}

Finally, when taking the infinitely thin or WLC limit for long chains
($L\gg b$), we can assume that around $\mathbf{r}=0$, $\mathbf{C}\left(\mathbf{r},\omega\right)\approx\frac{\mathbf{C}\left(0\right)}{4\pi}$
is approximately uniform and independent of $\omega$. Defining $\delta\mathbf{r}$
by $d_{min}=0$ and $\delta\omega'=4\pi$, while taking $\delta\mathbf{r}\rightarrow0$,
the expression above reduces to
\begin{equation}
J=\frac{4\pi\int\limits _{\delta\mathbf{r}}\int\limits _{\delta\omega}C\left(\mathbf{r},\omega\right)\mbox{d}\mathbf{r}\mbox{d}\omega}{N_{A}\sigma_{R_{N}}\delta\mathbf{r}\delta\omega}\approx\frac{\mathbf{C}\left(0\right)\delta\mathbf{r}\delta\omega}{N_{A}\sigma_{R_{N}}\delta\mathbf{r}\delta\omega}=\frac{\mathbf{C}\left(0\right)}{N_{A}\sigma_{R_{N}}},
\end{equation}
which is the original expression by Flory \cite{flory_statistical_1969}
for the WLC J-factor. Note that for the case of DNA cyclization $\sigma_{R_{N}}=1$
\cite{semlyen_cyclic_2000}. We use the expression in Eq.~(\ref{eq:J_factor})
to compute the J-factor by numerical simulation.

\section{\label{sec:numerical-implementation}Numerical Implementation of SAWLC}

\subsection{Numerical implementation overview}

%In this section we describe how to iteratively
%generate an ensemble with partition function that approaches (\ref{})
%as the number of generated chains approaches infinity.

In order to evaluate the SAWLC looping probabilities, we developed
a Monte-Carlo algorithm that is capable of generating a large number
of plausible self-avoiding DNA chains.
The Monte-Carlo simulation was written in CUDA C++, and was executed
on two NVIDIA GeForce GTX TITAN cards. The length of the generated chains ranged from $50$ to $5000$ links. Each chain in our ensemble was grown one link at a time by selecting the link's orientation according to the distribution described in Sec.~\ref{sub:sampling-chain-angles}, taking into account disallowed directions due to the excluded volume of previous links. In addition, each completed chain was assigned a Rosenbluth weight as described in Sec.~\ref{sup:faithful-ensemble-sampling}, resulting in a faithful representation of the full configurational space. Typical sizes of generated ensembles were $N_c\approx 10^{10}$ chains.

We computed the mean-square end-to-end distance
$\left\langle R^{2}\right\rangle $ and the polymer end-to-end separation
power-law exponent $\nu$ ($\sqrt{\left\langle R^{2}\right\rangle }\propto\mathit{N^{\nu}}$) in order to compare our simulation
to previous results \cite{clisby_accurate_2010, chen_renormalization-group_1992}. For the computation of the J-factor,
we computed the function $\mathbf{C}\left(\mathbf{r},\omega\right)$ describing the distribution of
the end-to-end vector $\mathbf{r}$ with angle $\omega$ from $\hat{t}_{1}$
(see Eq.~\ref{eq:J_factor}).
The procedures used in the computation
of these expressions are detailed in Sec.~\ref{sup:faithful-ensemble-sampling}.

In the following, we denote the number of links in the
simulated chain by $N$, the length of the links by $l$, the length
of the chain by $L\equiv Nl$, the diameter (or width) of the chain by
$w$ and the Kuhn length of the corresponding WLC as $b$, where $b$
is given by \cite{tree_is_2013}:
\begin{equation}
\frac{b}{l}=\left(\frac{a-1+a\coth a}{a+1-a\coth a}\right),
\end{equation}
and where $a$ is the bending constant of the chain.
%Unless stated otherwise, the values To simulate double-stranded DNA we used %the values of $b\approx106nm$
%for the Kuhn length of the DNA and $w\approx4.6nm$ for the width
%of the DNA.

\subsection{\label{sub:sampling-chain-angles}Sampling of chain angles}

Our goal was to sample angles $\theta_{i},\phi_{i}$ that satisfy the probability distribution
\begin{multline}
P_{i}^{\theta,\phi}\left(\cos\theta_{i}\epsilon\left[-1,1\right];\,\phi_{i}\epsilon[0,2\pi)\right)\propto\\
\exp\left[-\beta E^{el}\left(\theta_{i},\phi_{i}\right)\right]\Theta_{i}^{hw}\left(\left\{ \theta_{i},\phi_{i}\right\} \right)=\\
\exp\left[a\left(1-\cos\theta_{i}\right)\right]\Theta_{i}^{hw}\left(\left\{ \theta_{i},\phi_{i}\right\} \right)
\end{multline}\.
Note that we sampled $\cos\theta_{i}$ and
not $\theta_{i}$ according to this distribution, since the quantity that
is distributed uniformly over the unit sphere is $\cos\theta_{i}$
and not $\theta_{i}$.

We chose to sample the $i$th link's orientation angles $\left(\cos\theta_{i},\phi_{i}\right)$
using inversion sampling \cite{devroye_non-uniform_1986}. Here the
inversion sampling of a single random variable $X$ from a probability
distribution function ($PDF\left(X\right)$) is carried out by, first,
integrating the $PDF\left(X\right)$ over the entire range of $X$,
resulting in a cumulative distribution function ($CDF\left(X\right)$).
Then, a random number $y$ in the range $\left(0,1\right)$ is generated
from a continuous uniform distribution. Finally, the desired variable
value $x$ is extracted from the inverse function of the $CDF\left(X\right)$
at y, i.e. $x=CDF^{-1}\left(y\right)$. In the case discussed here,
the two-dimensional variable space $\left(\cos\theta_{i},\phi_{i}\right)$
can be mapped to a one-dimensional space due to azimuthal symmetry
considerations in the elastic energy. We computed the $CDF\left(X\right)$ numerically, using the adaptive Gaussian integration method, \cite{pieper_recursive_1999} and subsequently
inverted it.

In order to show how the mapping from the two-dimensional variable
space $\left(\cos\theta_{i},\phi_{i}\right)$ to a one-dimensional
space is carried out, we first consider the case of the chain without
the excluded volume constraint. Namely, we compute the inverse of
the following function:
\begin{equation}
I\left(\cos\theta_{i}\right)=\frac{\int\limits _{-1}^{\cos\theta_{i}}\mbox{d}\cos\theta\int\limits _{0}^{2\pi}\mbox{d}\phi\, e^{-a\left(1-\cos\theta\right)}}{\int\limits _{-1}^{1}\mbox{d}\cos\theta\int\limits _{0}^{2\pi}\mbox{d}\phi\, e^{-a\left(1-\cos\theta\right)}},
\end{equation}
which allows us to extract $\cos\theta_{i}$, provided that some random
number $y$ corresponding to $I\left(\cos\theta_{i}\right)$ is generated in the range $\left(0,1\right)$. Finally, we generate $\phi_{i}$
from a uniform distribution over the entire range $\phi_{i}\epsilon[0,2\pi)$
to complete the two-angle set.

If we reintroduce the hard-wall potential, the above integral now
changes to:
\begin{equation}
I\left(\cos\theta_{i}\right)=\frac{\int\limits _{-1}^{\cos\theta_{i}}\mbox{d}\cos\theta\int\limits _{allowed\,\phi\, for\,\theta}\mbox{d}\phi\, e^{-a\left(1-\cos\theta\right)}}{\int\limits _{-1}^{1}\mbox{d}\cos\theta\int\limits _{allowed\,\phi\, for\,\theta}\mbox{d}\phi\, e^{-a\left(1-\cos\theta\right)}}.\label{eq:SUP-MC_sampling_int}
\end{equation}

\begin{figure}
\includegraphics[width=1\columnwidth]{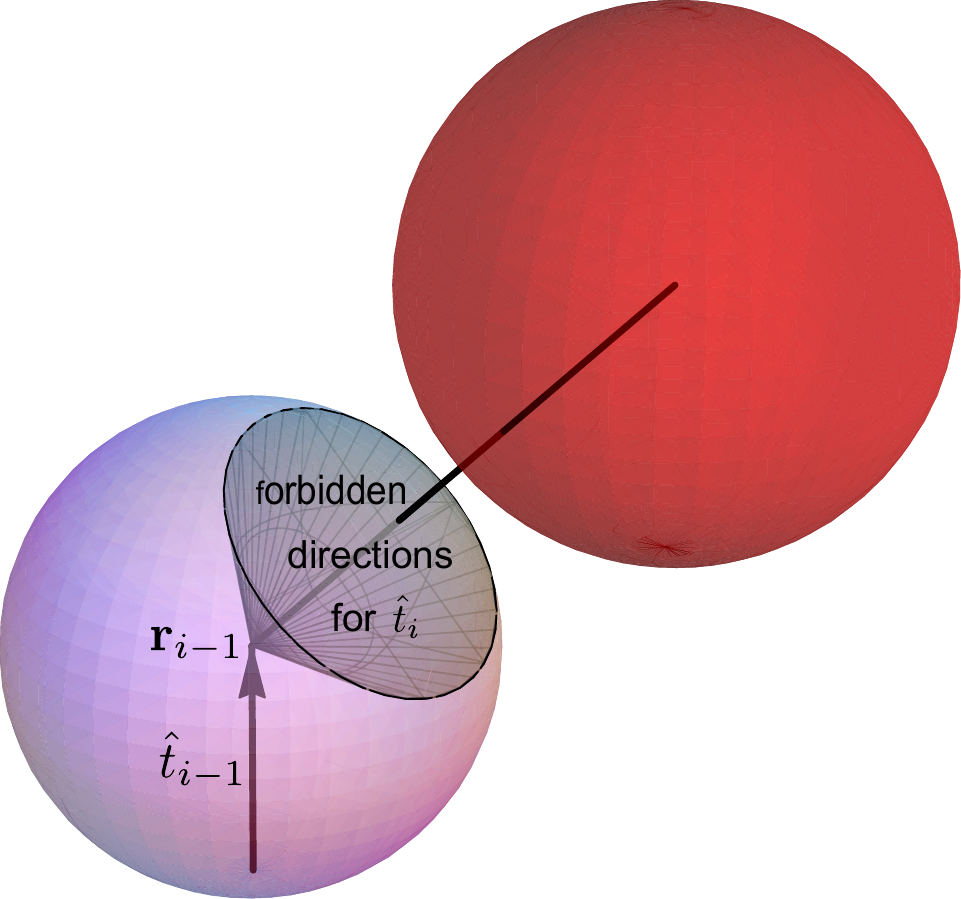}
\caption{\label{fig:SUP-Forbidden_ranges}(Color online) Forbidden directions
for $\hat{t}_{i}$. Purple (light gray) sphere is the excluded
volume around $\mathbf{r}_{i-1}$ (the endpoint of the $(i-1)$th link).
The red (dark gray) sphere is a potentially colliding object. If $\hat{t}_{i}$
assumes any direction inside the black (dark gray) cone this would result in a
collision of the hard-wall sphere at $\mathbf{r}_{i}$ with the red
sphere.}
\end{figure}

The integral over $\phi$ is no longer over the entire
$[0,2\pi)$ range, as some directions in space cannot be assumed by
the new link due to the excluded volume constraint. Thus, for each
possible $\theta$ angle of the new link, there is a different range
of $\phi$ values that are allowed. The situation is illustrated in
Fig.~\ref{fig:SUP-Forbidden_ranges}. The red sphere is an object
in close proximity to the endpoint of the $\left(i-1\right)$th link.
The black cone maps the directions forbidden for $\hat{t}_{i}$, as
that would cause an overlap between the $i$th link and the red sphere.
These directions must be excluded from the integration in (\ref{eq:SUP-MC_sampling_int}). As a result, we may proceed similarly to the non-constrained case,
except that in this case the angle $\phi_{i}$ is no longer sampled
from the full $[0,2\pi)$ range, but from a smaller range of possible
values.

In order to compute the integral in (\ref{eq:SUP-MC_sampling_int})
and to generate $\phi_{i}$, we first compute the set of values that
are allowed for $\phi_{i}$ for each value of $\theta_{i}$. Since
there can be more than one colliding object, the set of allowed values
for $\phi_{i}$ need not be one continuous range. Rather, these values
can be organized into $m\ge1$ consecutive ranges $[r_{i,0}^{a},r_{i,0}^{b}],...,[r_{i,m-1}^{a},r_{i,m-1}^{b}]$,
such that
\begin{equation}
\phi{}_{i}^{tot}=\sum_{j=0}^{m-1}\left(r_{i,j}^{b}-r_{i,j}^{a}\right)\leq2\pi.
\end{equation}

Thus, for each $\theta_{i}$ the $\phi_{i}$ angle can be sampled
from a stepwise uniform distribution:
\begin{equation}
P_{i}^{\phi}\left(\theta_{i},\,\phi_{i}\epsilon(0,2\pi)\right)=\frac{\sum\limits _{j=0}^{m-1}\Theta\left(\phi_{i}-r_{i,j}^{a}\right)\Theta\left(r_{i,j}^{b}-\phi_{i}\right)}{\phi{}_{i}^{tot}}.
\end{equation}

\subsubsection{\label{sub:Mapping-disallowed-directions}Mapping disallowed directions for the $i$th link}

After generating $i-1$ links of the chain, the endpoint is located
at $\mathbf{r}_{i-1}$ and the link direction is $\hat{t}_{i-1}$.
We now have to map all the directions that $\hat{t}_{i}$ cannot assume
(as shown in Fig.~\ref{fig:SUP-Forbidden_ranges}). In the following
discussion we focus on one of the chain joints in the vicinity of
$\mathbf{r}_{i-1}$ which limits the range of $\hat{t}_{i}$. We denote
the location of this potentially colliding joint as $\mathbf{r}_{p}$
and the vector from the endpoint $\mathbf{r}_{i-1}$ to the center
of the potentially colliding sphere as $\mathbf{r}$:
\begin{equation}
\mathbf{r}=\mathbf{r}_{p}-\mathbf{r}_{i-1}.
\end{equation}
We denote by $\gamma$ the angle between the direction of the last
link $\hat{t}_{i-1}$ and $\mathbf{r}$:
\begin{equation}
\cos\gamma=\frac{\mathbf{r}}{\left|\mathbf{r}\right|}\cdot\hat{t}_{i-1}.
\end{equation}
We note that the forbidden directions for $\hat{t}_{i}$ form a cone
around $\mathbf{r}$ (Fig.~\ref{fig:SUP-Forbidden_ranges}). We
denote half of the opening angle of the cone by $\alpha$ which can
then be calculated from simple geometrical considerations as:
\begin{equation}
\cos\alpha=\frac{\left|\mathbf{r}\right|^{2}-w^{2}+l^{2}}{2\left|\mathbf{r}\right|l}.
\end{equation}
A certain $\left(\theta_{i},\phi_{i}\right)$ direction is disallowed
for $\hat{t}_{i}$ if the vector $\mathbf{r}_{i}=\mathbf{r}_{i-1}+l\hat{t}_{i}$
is at a distance $\left|\mathbf{r}-\mathbf{r}_{i}\right|^{2}\leq w^{2}$
from $\mathbf{r}_{p}$. We proceed to work in the local coordinates
of the $\left(i-1\right)$th link, assuming for now that $\mathbf{r}$
is in the XZ plane in these coordinates:
\begin{eqnarray}
\mathbf{r} & = & \left|\mathbf{r}\right|\left(\sin\gamma,0,\cos\gamma\right),\\
\mathbf{r}_{i}=l\hat{t}_{i} & = & l(\sin\theta_{i}\cos\phi_{i},\sin\theta_{i}\sin\phi_{i},\cos\theta_{i}).
\end{eqnarray}

Disallowed values of $\left(\theta_{i},\phi_{i}\right)$ satisfy the
following inequality:
\begin{multline}
\left|\mathbf{r}-\mathbf{r}_{i}\right|^{2}=(\left|\mathbf{r}\right|\sin\gamma-l\sin\theta_{i}\cos\phi_{i})^{2}+\\
(\left|\mathbf{r}\right|\cos\gamma-l\cos\theta_{i})^{2}+l^{2}\sin^{2}\theta_{i}\sin^{2}\phi_{i}\leq w^{2}.\label{eq:SUP-collision_detection_touch}
\end{multline}
We can thus map the disallowed range for $\hat{t}_{i}$ in two steps.
We first determine the range of $\cos\theta_{i}$ for which a collision
can occur at all. Then, for each $\theta_{i}$ in this range we determine
the range of $\phi_{i}$ for which a collision does occur.

\subsubsection{The $\theta_{i}$ range for which a collision occurs}

The critical values $\cos\theta_{\pm}$ for which $\left|\mathbf{r}-\mathbf{r}_{i}\right|^{2}=w^{2}$
can be derived from Eq.~(\ref{eq:SUP-collision_detection_touch}):
\begin{equation}
\cos\theta_{\pm}=\cos\left(\gamma\mp\alpha\right)=\cos\gamma\cos\alpha\pm\sin\gamma\sin\alpha.\label{eq:SUP-disallowed_theta}
\end{equation}

\begin{figure}[thb]
\includegraphics[width=1\columnwidth]{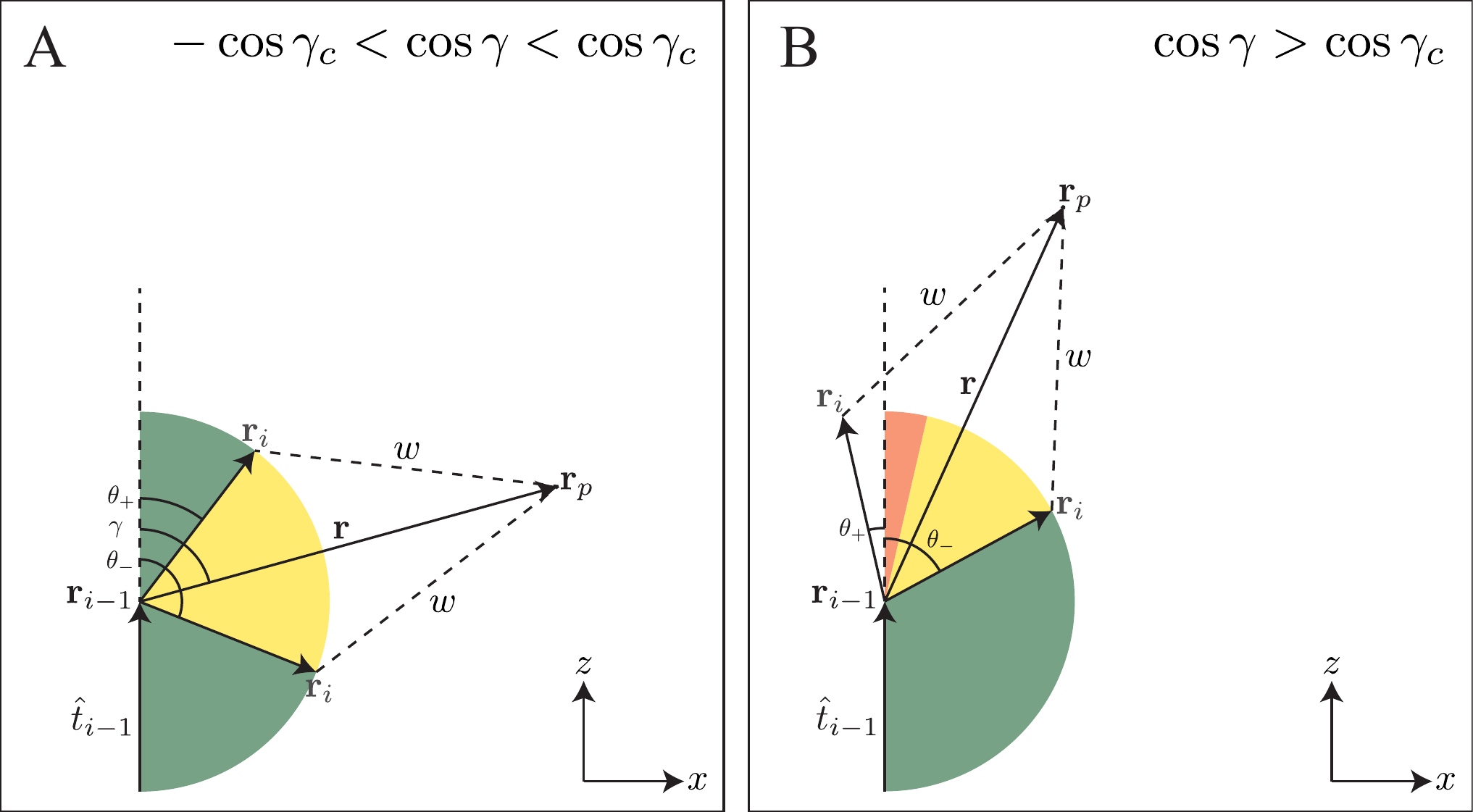}
\caption{\label{fig:SUP-forbidden_thetas}(Color online) $\theta_{i}$
ranges mapping. $\theta_{i}$ ranges that do not
cause a collision for all values of $\phi_{i}$ are marked in green (dark gray).
$\theta_{i}$ ranges that cause a collision for some values of $\phi_{i}$
are marked in yellow (light gray). $\theta_{i}$ ranges that cause a collision
for all values of $\phi_{i}$ are marked in red (medium gray). (a) If $\theta_{i}\in[\theta_{+},\theta_{-}]$,
a collision occurs for some values of $\phi_{i}$. (b) If $\theta_{i}\in[0,\theta_{+}]$,
a collision occurs for all values of $\phi_{i}$.}
\end{figure}

Caution is needed when interpreting the results in (\ref{eq:SUP-disallowed_theta}).
It is true that choosing $\cos\theta_{i}\in\left[\cos\theta_{-},\cos\theta_{+}\right]$
will result in collision for some values of $\phi_{i}$ (Fig.~\ref{fig:SUP-forbidden_thetas}).
However, this does not take into account additional $\theta_{i}$
ranges for which there is a collision for all values of $\phi_{i}$.
Such a situation is shown in Fig.~\ref{fig:SUP-forbidden_thetas}(b):
for $\cos\theta_{i}\in\left[\cos\theta_{-},\cos\theta_{+}\right]$
obviously a collision occurs for some values of $\phi_{i}$, but for
$\cos\theta_{i}\in\left[\cos\theta_{+},1\right]$ a collision occurs
for all values of $\phi_{i}$.

If the colliding object is located at $\cos\gamma>\cos\alpha$ or
at $\cos\gamma<-\cos\alpha$ there is a range of $\theta_{i}$ values
for which a collision occurs for all values of $\phi_{i}$. To summarize:
\begin{itemize}
\item There are disallowed $\phi_{i}$ angles for $\cos\theta_{i}\in[\cos\theta_{-},\cos\theta_{+}]$.
\item If $\cos\gamma>\cos\alpha$, for $\cos\theta_{i}\in[\cos\theta_{+},1]$
there is always a collision.
\item If $\cos\gamma<-\cos\alpha$, for $\cos\theta_{i}\in[-1,\cos\theta_{-}]$
there is always a collision.
\end{itemize}

\subsubsection{The $\phi_{i}$ range in which a collision occurs, for specific $\theta_{i}$}

We begin with Eq.~(\ref{eq:SUP-collision_detection_touch}) and find the
critical value of $\phi_{i}$ ($\phi_{ic}$) for a specific value
of $\theta_{i}$:
\begin{equation}
\cos\phi_{ic}\left(\theta_{i}\right)=\frac{\cos\alpha-\cos\gamma\cos\theta_{i}}{\sin\gamma\sin\theta_{i}}.\label{eq:SUP-disallowed_phi}
\end{equation}
A full collision (no possible values for $\phi_{ic}$) occurs when
\begin{eqnarray}
\frac{\cos\alpha-\cos\gamma\cos\theta_{i}}{\sin\gamma\sin\theta_{i}} & < & -1.
\end{eqnarray}
No collision occurs when
\begin{eqnarray}
\frac{\cos\alpha-\cos\gamma\cos\theta_{i}}{\sin\gamma\sin\theta_{i}} & > & 1.
\end{eqnarray}
The disallowed angles for a specific $\theta_{i}$ are:
\begin{equation}
-\phi_{ic}\left(\theta_{i}\right)\leq\phi_{disallowed}\left(\theta_{i}\right)\leq\phi_{ic}\left(\theta_{i}\right).
\end{equation}

We can now generalize to the case when $\mathbf{r}$ is not in the
XZ plane. If $\mathbf{r}=\left|\mathbf{r}\right|\left(\sin\gamma\cos\beta,\sin\gamma\sin\beta,\cos\gamma\right)$
then
\begin{equation}
\beta-\phi_{ic}\left(\theta_{i}\right)\leq\phi_{disallowed}\left(\theta_{i}\right)\leq\beta+\phi_{ic}\left(\theta_{i}\right).
\end{equation}

\subsection{\label{sup:faithful-ensemble-sampling}Faithful ensemble sampling}

For the simulation of the WLC model we used importance sampling (IS,
see Appendix~\ref{sub:SUP-Importance-sampling}). For the simulation of the SAWLC model (and the thick freely-jointed
self-avoiding chain) we used weighted-biased sampling (WBS, see Appendix~\ref{sub:SUP-weighted-biased-sampling})
\cite{prigogine_monte_1998}, which is based on a method developed
by Rosenbluth and Rosenbluth \cite{rosenbluth_monte_1955}. A variation of this method was recently
used to sample dsDNA configurational space \cite{tree_is_2013} for
much longer chains than the ones used in this work to examine the
cyclization process.

The partition function for an ensemble of $N_{c}$ sampled chains
of $N$ links in WBS is given by
\begin{equation}
Z_{N}=\sum_{j=1}^{N_{c}}W\left(\left\{ \theta_{N},\phi_{N}\right\} _{j}\right),
\end{equation}
where $W\left(\left\{ \theta_{N},\phi_{N}\right\} _{j}\right)$ is
the Rosenbluth factor of the $j$th chain in the ensemble:
\begin{widetext}
\begin{equation}
W\left(\left\{ \theta_{N},\phi_{N}\right\} \right)  =\prod_{i=1}^{N}  \int\limits _{-1}^{1}\mbox{d}\cos\theta_{i}\int\limits _{0}^{2\pi}\mbox{d}\phi_{i}\exp\left[-\beta E^{el}\left(\theta_{i},\phi_{i}\right)\right]\Theta_{i}^{hw}\left(\left\{ \theta_{i},\phi_{i}\right\} \right) =  \prod_{i=1}^{N} \!\!\!\!\!\!\!\!\!\!\!\!\!\!\!\!\!\!\int\limits _{non-colliding\, directions}\!\!\!\!\!\!\!\!\!\!\!\!\!\!\!\!\!\!\mbox{d}\theta_{i}\mbox{d}\phi_{i}\exp\left[-\beta E^{el}\left(\theta_{i},\phi_{i}\right)\right],\label{eq:W}
\end{equation}
\end{widetext}
with $\Theta_{i}^{hw}$ defined in eq.~(\ref{eq:Theta-HW}).

The average value $\left\langle f\right\rangle $ of a physical property
is calculated as:
\begin{equation}
\left\langle f\right\rangle =\frac{\sum\limits _{j=1}^{N_{c}}f\left(\left\{ \theta_{N},\phi_{N}\right\} _{j}\right)W\left(\left\{ \theta_{N},\phi_{N}\right\} _{j}\right)}{\sum\limits _{j=1}^{N_{c}}W\left(\left\{ \theta_{N},\phi_{N}\right\} _{j}\right)}.\label{eq:MC_<f>}
\end{equation}
\\
This allows the computation of $\left\langle R^{2}\right\rangle$. The power-law exponent $\nu$ was extracted from the computed $\left\langle R^{2}\right\rangle$ (see Appendix~\ref{sub:SUP-power-law-exponent}).

To approximate $C\left(\mathbf{r},\omega\right)$, the configurational
space for $\mathbf{r}=\left(r,\omega'\right)$ and $\omega$ is partitioned
into bins, with bin volume $\mbox{d}\mathbf{r}\times\mbox{d}\omega$.
In each bin we compute the probability of the end-to-end vector $\mathbf{r}$
and relative bond orientation to be contained in the bin:
\begin{widetext}
\begin{equation}
P_{bin}\left(r,\omega',\omega\right)=\lim_{N_{c}\rightarrow\infty}\frac{\sum\limits _{j=1}^{N_{c}}f_{bin}\left(\mathbf{r},\omega,\left\{ \theta_{N},\phi_{N}\right\} _{j}\right)W\left(\left\{ \theta_{N},\phi_{N}\right\} _{j}\right)}{\sum\limits _{j=1}^{N_{c}}W\left(\left\{ \theta_{N},\phi_{N}\right\} _{j}\right)}
\end{equation}
where
\begin{eqnarray}
f_{bin}\left(\mathbf{r},\omega,\left\{ \theta_{N},\phi_{N}\right\} \right) & = & \begin{cases}
1 & \,\,\,\,\mbox{\ensuremath{\mathbf{r}_{N}} and \ensuremath{\hat{t}_{N}} relative to \ensuremath{\hat{t}_{1}} are contained}\\
 & \,\,\,\,\mbox{ in the bin defined by \ensuremath{\mathbf{r}},\ensuremath{\omega}\ and \ensuremath{\mathbf{\mbox{d}r}},d\ensuremath{\omega}}\\
0 & \,\,\,\,\mbox{otherwise}
\end{cases}.
\end{eqnarray}
\end{widetext}
$C\left(\mathbf{r},\omega\right)$ is then computed by
\begin{equation}
C\left(\mathbf{r},\omega\right)=\frac{P_{bin}\left(r,\omega',\omega\right)}{\mbox{d}\mathbf{r}\times\mbox{d}\omega}.
\end{equation}

\section{\label{sec:results}Results}

\subsection{\label{sub:Results-DNA-end-to-end-distance}DNA end-to-end distance analysis and comparison to RG results}

In order to verify that our algorithm generates a faithful swollen
chain ensemble, we simulated chains of up to 5000 links and compared
our simulation results to previously-published renormalization group
(RG) calculations \cite{chen_renormalization-group_1992} and Monte-Carlo
results \cite{tree_is_2013}. We simulated chains with the touching-beads
model ($l=w$, see Fig.~\ref{fig:WLC-SAWLC}(b)) and dsDNA-like parameters
of $b\approx106\mbox{ nm}$ and $w\approx4.6\mbox{ nm}$.

\begin{figure}
\includegraphics[width=1\columnwidth]{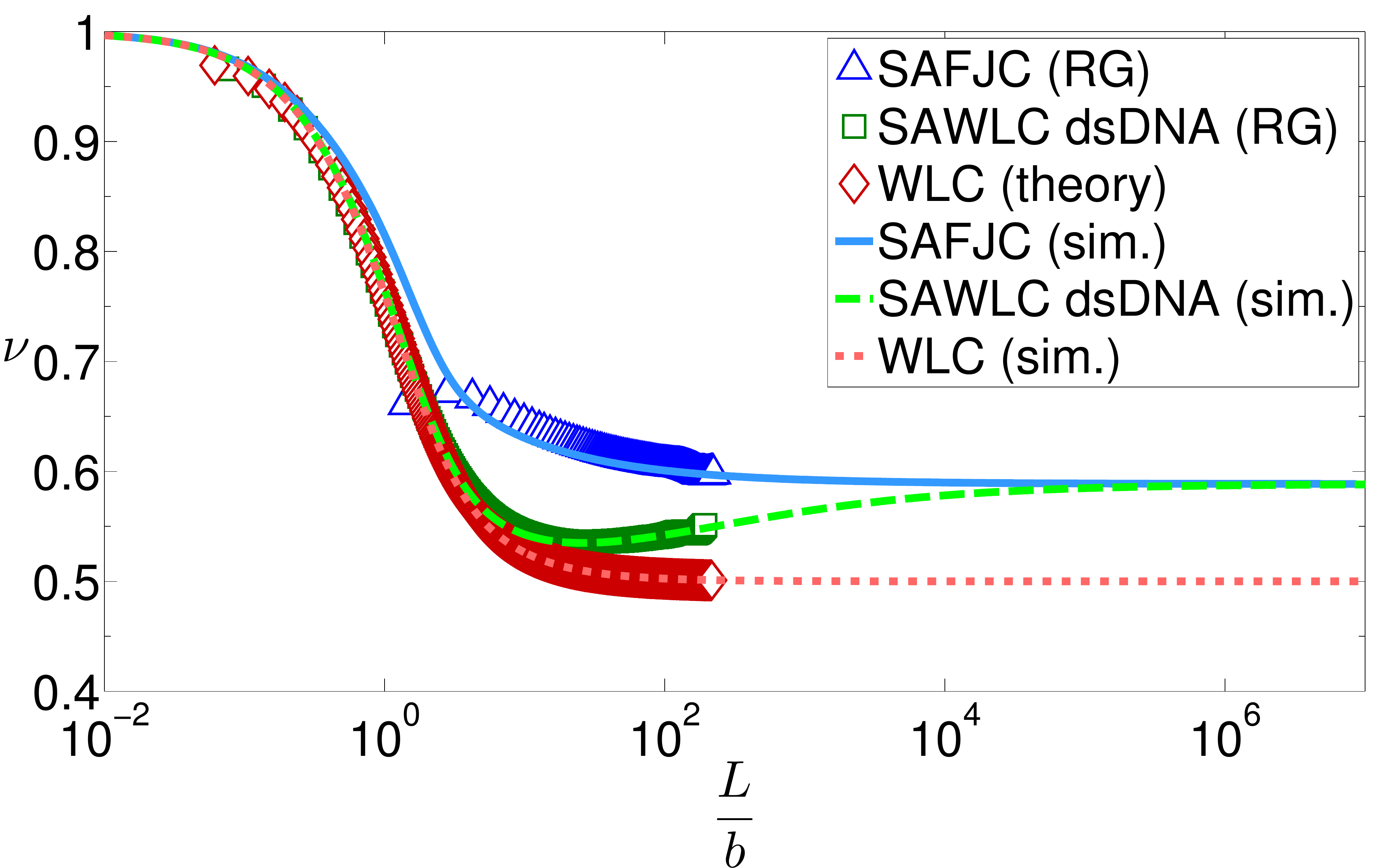}
\caption{\label{fig:Tree_Results}(Color online) Comparison of the simulation results
to the RG calculations and WLC theory. Power-law exponent
$\nu$ of the end-to-end distance $\sqrt{\langle R^{2}\rangle}$ is plotted as a function of chain
length $L/b$. We compare three cases (top to bottom): a feely-jointed
``thick'' polymer (SAFJC) with $\frac{w}{b}=1$ and $l=w$ (solid blue line -
RG and blue triangles - simulation), a ``thick'' polymer (SAWLC) with dsDNA-like parameters $\frac{w}{b}=\frac{4.6}{106}\ll1$
and $l=w$ (dashed green line - RG and green squares - simulation),
and a zero-thickness dsDNA (WLC) with $\frac{l}{b}=\frac{4.6}{106}$
(dotted red line - WLC and red diamonds - simulation). The excluded volume
parameter in the RG calculations was rescaled to match the excluded
volume in the simulations with $\overline{u}_{RG}=1.5u$ following
Tree et. al.~\cite{tree_is_2013}.}
\end{figure}

In Fig.~\ref{fig:Tree_Results} we plot the RG predictions and our
simulation results for the polymer end-to-end separation power-law
exponent $\nu$ as a function of the normalized chain length $\frac{L}{b}$. We display three cases: a ``thick'' polymer with
dsDNA-like parameters $\frac{w}{b}\approx\frac{4.6}{106}\ll1$ (SAWLC),
a zero-thickness dsDNA (WLC) and a fully flexible ``thick'' polymer
with $\frac{w}{b}=1$ (self-avoiding freely-jointed chain, SAFJC). The simulation
shows that the SAWLC (dashed green curve and green squares) maintains
a WLC-like behavior (dotted red curve and red diamonds) for lengths that
are up to several times larger than the persistence length $l_{P}\equiv\frac{b}{2}$.
At $\frac{L}{b}\approx30$ the curve and simulation begin to bend
up and approach the SAFJC prediction until convergence of
the SAWLC and the SAFJC is predicted
to occur by RG at $\frac{L}{b}\approx10^{4}$. It is important to
note that while our simulation does not capture the exact convergence
to the predicted SAWLC behavior due to chain-length limitations
of our algorithm, the trends are precisely tracked over the range
of the RG predictions. Moreover, the length at which the divergence
away from the WLC behavior occurs (the length where the dashed green and dotted red
lines separate) is highly dependent on the width parameter $\left(\frac{w}{b}\right)$. This result is consistent
with experimental observations \cite{nepal_structure_2013} and simulation
\cite{tree_is_2013}.

For the case of the SAFJC $\left(\frac{w}{b}=1\right)$,
our simulation produces a decaying power law that converges to the
Flory prediction of $\sqrt{\langle R^{2}\rangle}\propto N^{-0.5876}$
for long chain lengths. Note that our simulation for the
semi-flexible chain deviates from
the RG predictions only for very short chains ($N<5$). However, for the fully flexible chain
our simulation converges on the RG predictions for values $N>20$. These deviations are a natural consequence of the discrete
simulation process, as our simulation employs finite size links while
the RG and WLC predictions are for continuous chain models, which
correspond to the case $l\rightarrow0$ and $N\rightarrow\infty$.
The larger short-scale deviation from the RG predictions generated
by our simulation for the SAFJC (deviation of the blue triangles
from the solid blue line for $\frac{L}{b}<10^{2}$) can be explained
by short-range excluded volume interactions, which are properly simulated
by our algorithm and are for the most part neglected in the RG model
\cite{chen_renormalization-group_1992}.

\subsection{\label{sub:J-entropic}The SAWLC J-factor in the entropic
regime}

\begin{figure*}
\includegraphics[width=2.075\columnwidth]{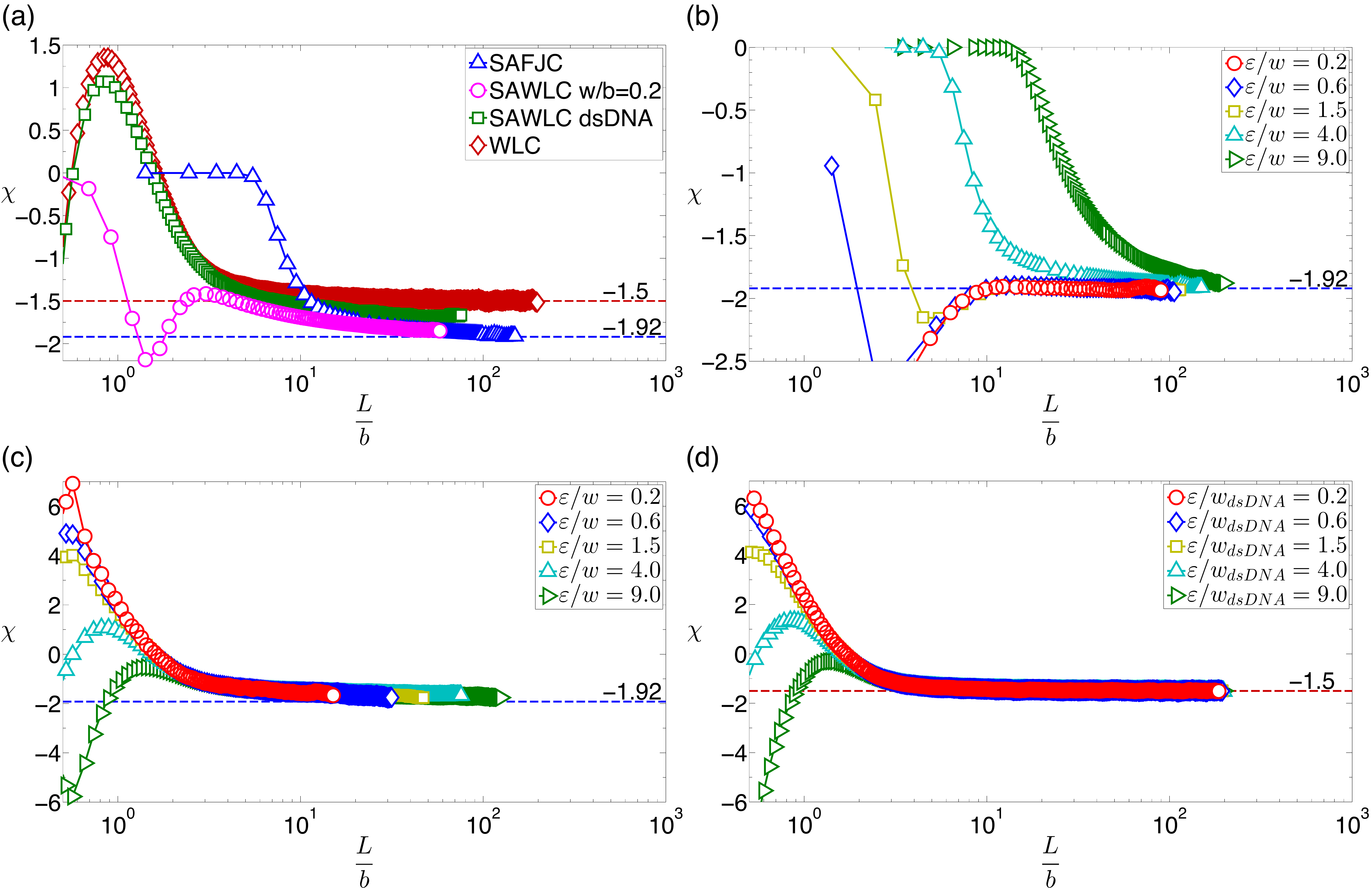}
\caption{\label{fig:Jfactor-power-law}(Color online)
The J-factor power law.
(a) Comparison of J-factor power-law exponents $\chi$ for the WLC ($\frac{w}{b}=0$) (red diamonds), SAWLCs with $\frac{w}{b}=0.046$ (green squares) and $\frac{w}{b}=0.2$ (magenta circles), and the SAFJC ($\frac{w}{b}=1$) (blue triangles), using $d_{min}=w$ and $\frac{\varepsilon}{w}=4$.
The blue(bottom) and red (top) dashed lines correspond to the SAFJC and WLC power laws of $-1.92$ and $-1.5$, respectively. (b)-(d) Effect of $\varepsilon$
on the J-factor power-law exponent. (b) J-factor power-law exponent for the SAFJC with various $\frac{\varepsilon}{w}$. (c) J-factor power-law exponent for the SAWLC with $\frac{\omega}{b}=\frac{4.6}{106}$ and with various $\frac{\varepsilon}{w}$. (d) J-factor power-law exponent for the WLC ($\frac{w}{b}=0$) with $b=106 $~nm and with various $\frac{\varepsilon}{w_{dsDNA}}$, where $w_{dsDNA}$ is taken to be equal to that of dsDNA.}
\end{figure*}

We next proceeded to model the looping or cyclization J-factor as
defined in Eq.~(\ref{eq:J_factor}). We first examined the entropic regime,
namely $\frac{L}{b}\gg1$. For the sake of simplicity, for this regime
we set $d_{min}=w$ and $\delta\omega'=\delta\omega=4\pi$ in (\ref{enu:bond-formation-1})
and (\ref{enu:bond-formation-2}). In Fig.~\ref{fig:Jfactor-power-law}
we plot the J-factor power-law exponent $\chi$ as a function of the
normalized chain length $\frac{L}{b}$. Previous theoretical studies
have shown that in the entropic regime, the J-factor scales like $J\propto N^{\chi}=N^{-\frac{3}{2}}$
for the WLC model and $J\propto N^{-\left(3\nu+\gamma-1\right)}=N^{-1.9196}$
for a flexible swollen chain (see Appendix~\ref{sub:SUP-asymptotic-J-power-law-derivation}),
where $\nu$ is the end-to-end power-law exponent and $\gamma$ is
the ``universal'' exponent \cite{gennes_scaling_1979}. In Fig.~\ref{fig:Jfactor-power-law}(a), we plot the J-factor power-law
exponents for the WLC ($\frac{w}{b}=0$, red diamonds), SAWLCs with $\frac{w}{b}=\frac{4.6}{106}$ (green squares) and $\frac{w}{b}=0.2$ (magenta circles), and the SAFJC ($\frac{w}{b}=1$, blue triangles). The panel
shows that the behavior of the power-law exponent dramatically changes
as the value of $\frac{w}{b}$ increases from zero for the ideal WLC
behavior to one for the fully swollen chain. Both the semi-flexible
WLC and the fully-flexible SAWLC J-factors converge
on their respective theoretical predictions. Moreover, our calculations
indicate that the SAWLC J-factor converges to the asymptotic value of $-1.92$ in a manner which is highly
dependent on the thickness of the respective chains.

To gain a better insight into the power-law behavior of the J-factor,
we plot in Fig.~\ref{fig:Jfactor-power-law}(b)-(d) the J-factor
power-law exponents for the SAFJC, the SAWLC with dsDNA aspect ratio $\frac{w}{b}=\frac{4.6}{106}$, and
the WLC, for various values of $\frac{\varepsilon}{w}$,
corresponding to the size of the sphere where bond formation between
the two chain termini occurs (see Eq.~\ref{enu:bond-formation-1}) divided
by the chain diameter. We notice that in all three figures the power-law
behavior in the far entropic regime converges to the scaling theory's
predicted limit. Instead, the effects of $\frac{\varepsilon}{w}$
seem to be localized to the shorter range elastic regime and the intermediate
(elastic-entropic) transition region. However, the effects of varying
$\frac{\varepsilon}{w}$ on the J-factor power-law exponent for the SAFJC are striking as compared with the other cases. Here
(Fig.~\ref{fig:Jfactor-power-law}(b)) we observe a sharp dependence
in the elastic regime. Moreover, the value of $\frac{L}{b}$ at which
the transition to the entropic regime and the convergence to the scaling-theory
power-law exponent occur, is a strong function of $\frac{\varepsilon}{w}$.
Consequently, we conclude that this parameter plays an important role
in the physical properties of polymers the closer the value of $\mathit{\frac{w}{b}}$
is to one.

\subsection{The SAWLC J-factor in the elastic regime}

\begin{figure}
\centering{}\includegraphics[width=1\columnwidth]{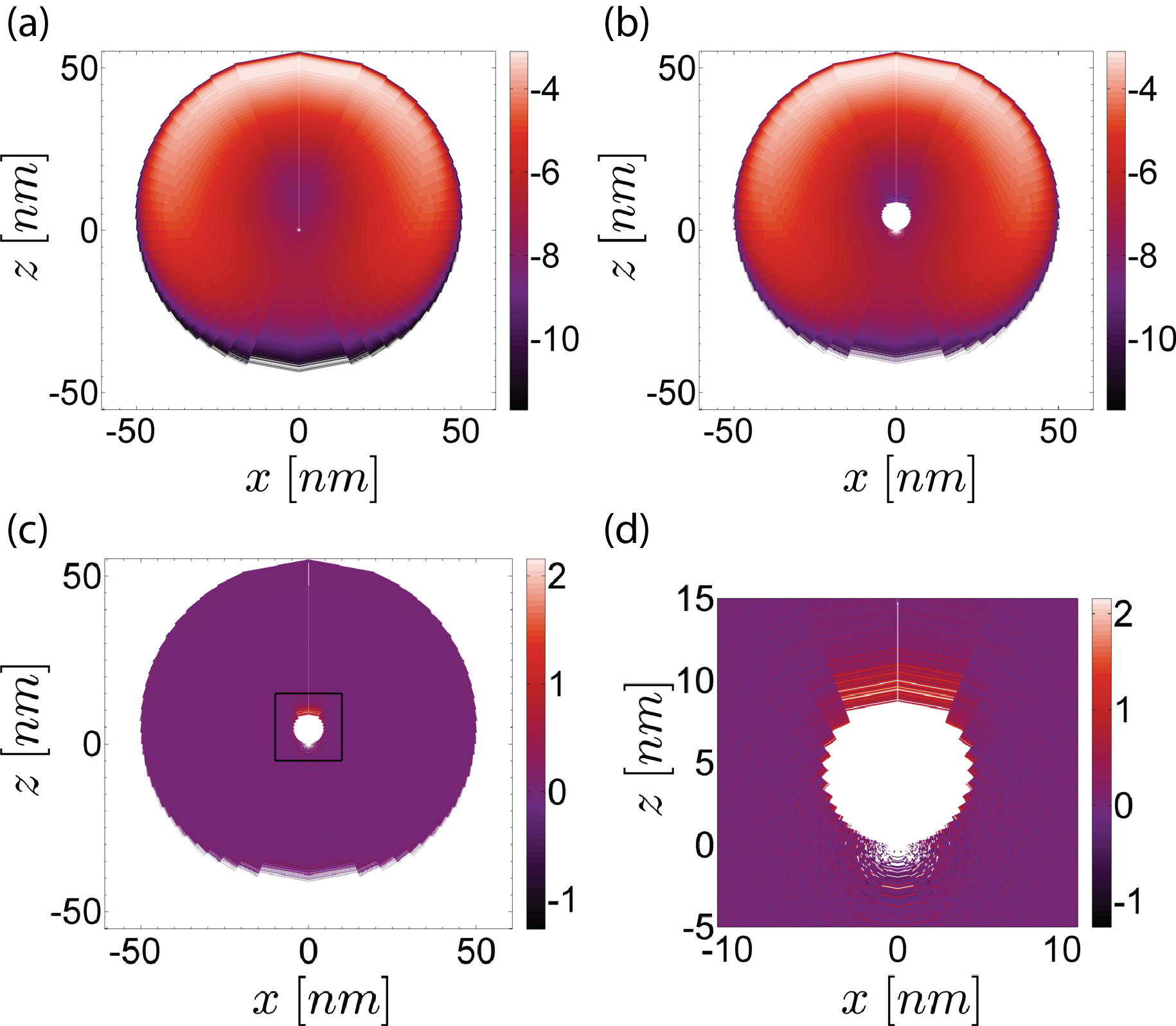}
\caption{\label{fig:WLC-and-SAWLCdmin0-comparison}(Color online) WLC and SAWLC ($d_{min}=0$) comparison in the elastic regime. The insets show a cross section of the data in the xz plane. $\mathbf{r}_{0}$
is in the origin and $\hat{t}_{1}=\hat{z}$. (a) $\log_{10}\left(\mathbf{C}_{\mathrm{WLC}}\mathbf{\left(\mathbf{r}\right)}/N_{A}\right)$
in the xz plane for WLC with $L=162$~bp. (b) $\log_{10}\left(\mathbf{C}_{\mathrm{SAWLC-0}}\mathbf{\left(\mathbf{r}\right)}/N_{A}\right)$
in the xz plane for SAWLC with $L=162$~bp, $\frac{w}{b}=\frac{4.6}{106}$
and $d_{min}=0$. (c) $\log_{10}\left(\frac{\mathbf{C}_{\mathrm{WLC}}\mathbf{\left(\mathbf{r}\right)}}{\mathbf{C}_{\mathrm{SAWLC-0}}\mathbf{\left(\mathbf{r}\right)}}\right)$.
(d) A blowup of the rectangular section marked in (c). Note,
shorter chains were also examined, however, the accumulated statistical
data around the origin in those chains was insufficient, and precluded
a comprehensive analysis. Qualitatively, however, the shorter chains
exhibited the same behavior as shown below in the data ranges that
contained significant statistical data.}
\end{figure}

In order to complete our description of the effects of the excluded
volume on the J-factor, we generated ensembles of short chains up to
$\frac{L}{b}\approx3$. It was previously predicted \cite{gennes_scaling_1979}
that for short chains the excluded volume effect should be negligible,
as the elastic energy is expected to be the dominant contribution
to the probability of looping. Fig.~\ref{fig:WLC-and-SAWLCdmin0-comparison}
shows cross-sections of $\mathbf{C}\left(\mathbf{r}\right)\equiv\int\mathbf{C}\left(\mathbf{r},\omega\right)\mbox{d}\omega$
in the $\mathrm{xz}$ plane for both the WLC and the SAWLC with $\frac{L}{b}\approx\frac{55.2~\mathrm{nm}}{106~\mathrm{nm}}\approx0.52$
and $d_{min}=0$ and $\delta \omega '=  \delta \omega = 4\pi$ in the SAWLC case, with $\mathbf{r}_{0}$ at the origin and $\hat{t}_{1}=\hat{z}$. Fig.~\ref{fig:WLC-and-SAWLCdmin0-comparison}(a)-(b)
show that the spatial distributions for the end-to-end separations
are highly anisotropic functions in the z-direction for both the SAWLC and the WLC. This implies that the
probability of looping itself must be highly anisotropic. In addition, both distributions look
similar, where the only significant deviation observed for the self-avoiding
from the non-self-avoiding distribution emerges from the space occupied
by the chain around the origin (white circle in Fig.~\ref{fig:WLC-and-SAWLCdmin0-comparison}(b).
In order to quantify this observation, we divided the two distributions,
as shown in Fig.~\ref{fig:WLC-and-SAWLCdmin0-comparison}(c). It is
evident from this panel that while both the self-avoiding and non-self-avoiding $\mathbf{C}\left(\mathbf{r}\right)$ functions are similar,
they differ around the origin in the positive $\mathrm{z}$ direction,
where the SAWLC has reduced probability (Fig.~\ref{fig:WLC-and-SAWLCdmin0-comparison}(d)).
This is expected, since $\mathbf{r}_{1}=l\hat{z}$, rendering the volume
around the origin inaccessible for $\mathbf{r}_{N}$.

\begin{figure}
\includegraphics[width=1\columnwidth]{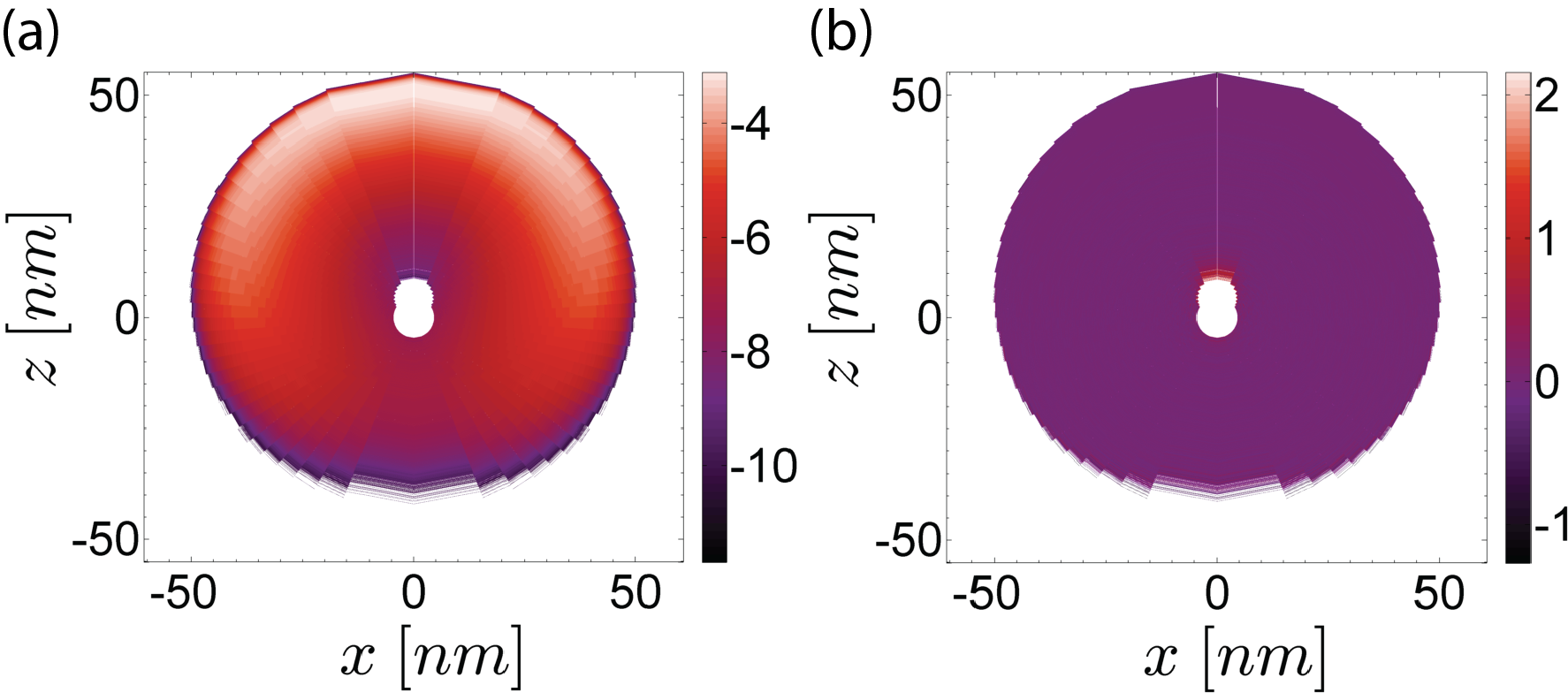}
\caption{\label{fig:SAWLC-SAWLCdmin0-comparison}(Color online)  WLC and SAWLC ($d_{min}=w$) comparison in the elastic regime. The insets show a cross section of the data in the xz plane. $\mathbf{r}_{0}$ is at the origin and $\hat{t}_{1}=\hat{z}$.
(a) $\log_{10}\left(\mathbf{C}_{\mathrm{SAWLC-w}}\mathbf{\left(\mathbf{r}\right)}/N_{A}\right)$
in the xz plane for SAWLC with $L=162$~bp, $\frac{w}{b}=\frac{4.6}{106}$
and $d_{min}=w$. (b) $\log_{10}\left(\frac{\mathbf{C}_{\mathrm{WLC}}\mathbf{\left(\mathbf{r}\right)}}{\mathbf{C}_{\mathrm{SAWLC-w}}\mathbf{\left(\mathbf{r}\right)}}\right)$.}
\end{figure}

We next examined the case in which $d_{min}\ne0$ (and $\delta \omega ' = \delta \omega = 4\pi$), implying that looping
is now defined for some finite volume around the origin for the WLC. In the SAWLC model, this condition translates to a larger
volume around the origin for which the distribution vanishes (see Fig.~\ref{fig:SAWLC-SAWLCdmin0-comparison}(a)). However, the actual deviation
from the non-self-avoiding WLC distribution function is for the most
part negligible, as can be seen from Fig.~\ref{fig:SAWLC-SAWLCdmin0-comparison}(b),
and is similar in its scope to the minimal effect described for $d_{min}=0$
above. Consequently, any deviation of the SAWLC J-factor from the WLC prediction in the elastic regime is not due to some global modification
of the underlying distribution, but rather arises from a local redefinition
of $\delta\mathbf{r}$ in Eq.~(\ref{eq:J_factor}), which is imperative
due to the presence of the excluded volume around the origin.

\begin{figure*}
\includegraphics[width=2.075\columnwidth]{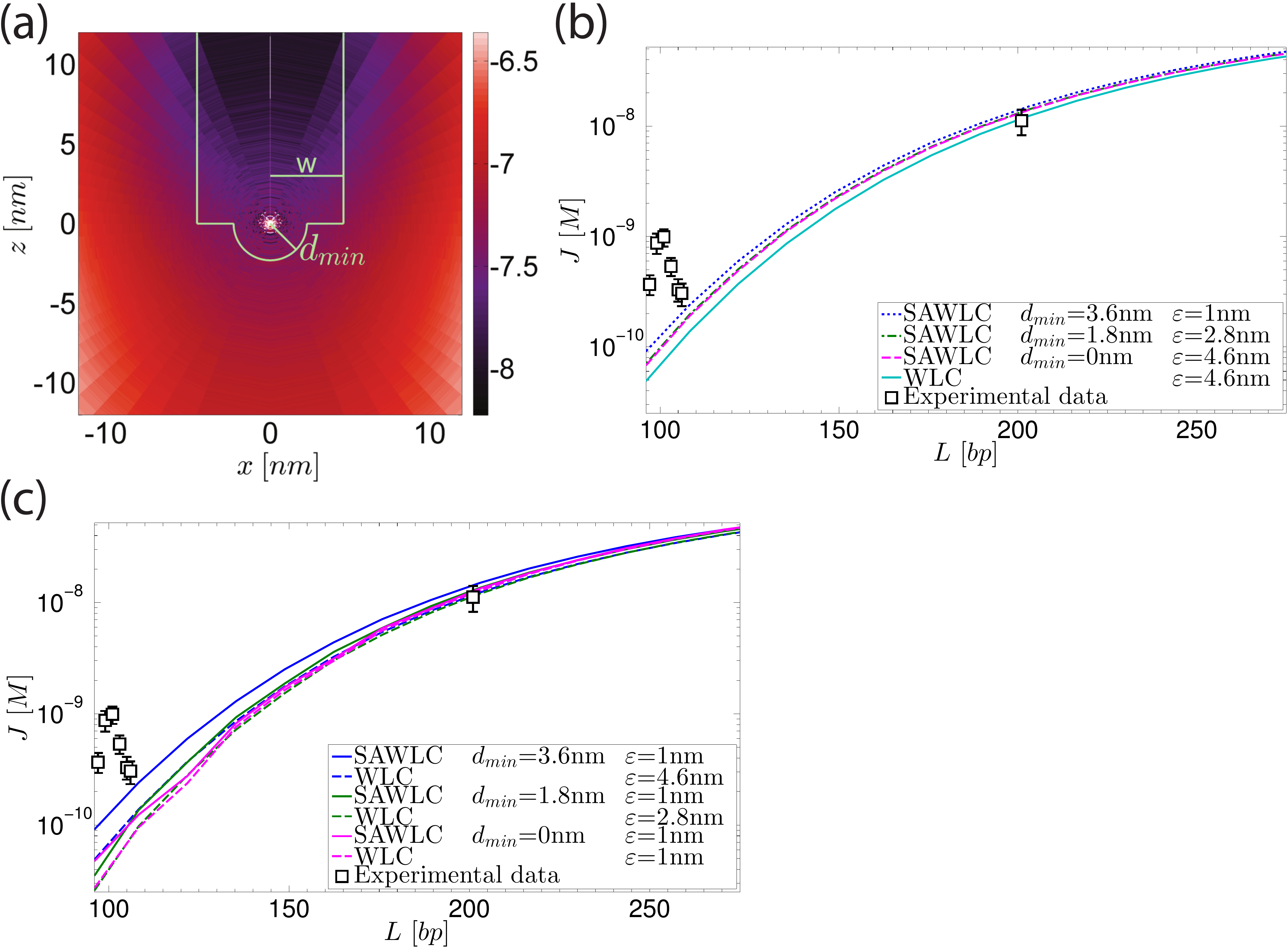}
\caption{\label{fig:WLC-removed-volume}(Color online) Effect of $d_{min}$ on the J-factor in the elastic regime. (a) The volume enclosed in the light-green outline is inaccessible to a SAWLC chain. (b) J-factor for the WLC and
SAWLC, varying $d_{min}$, constant $d_{min}+\varepsilon=4.6$~nm. Recent experimental data \cite{vafabakhsh_extreme_2012} is shown with black squares.
(c) J-factor for the WLC and SAWLC, varying $d_{min}$, constant $\varepsilon=1$~nm. Recent experimental data \cite{vafabakhsh_extreme_2012} is shown with black squares.
Note, the particular choice of $\mathit{d_{min}+\varepsilon}$ is
based on accepted physical values for the effective chain thickness
in standard saline conditions. Even though the effects are slightly
different in both cases, the magnitude of the effects is less than
half an order of magnitude for these values.}
\end{figure*}

Computation of $\mathbf{C}_{\mathrm{SAWLC}}(\mathbf{r})$ is extremely time-consuming for chain lengths shorter than $\approx 160$~bp. We therefore utilized the apparent identity of $\mathbf{C}_{\mathrm{SAWLC}}(\mathbf{r})$ and $\mathbf{C}_{\mathrm{WLC}}(\mathbf{r})$ far enough away from the origin to approximate $\mathbf{C}_{\mathrm{SAWLC}}(\mathbf{r})$ for shorter chain lengths. We modeled the effect of the excluded volume on the SAWLC J-factor
by computing $\mathbf{C}_{\mathrm{WLC}}\left(\mathbf{r}\right)$
and excluded volumes inaccessible to the SAWLC due to both $w>0$ and $d_{min}>0$ in the integration in Eq.~(\ref{eq:J_factor}). The removed volumes were approximated by a cylinder
with diameter $2w$ oriented along $\hat{z}$ with the center
of the lower base at the origin, and a sphere with radius $d_{min}$
centered at the origin. Fig.~\ref{fig:WLC-removed-volume}(a) shows
a schematic for the volume inaccessible to the SAWLC superimposed
on $\mathbf{C}_{\mathrm{WLC}}{\left(\mathbf{r}\right)}$.

Using this approximation for $\mathbf{C}_{\mathrm{SAWLC}}(\mathbf{r})$, we were able to
study the SAWLC J-factor for short chain lengths. In Fig.~\ref{fig:WLC-removed-volume}(b),
we study the case in which looping is defined for some constant bond-center
to bond-center separation ($d_{min}+\varepsilon=4.6$~nm) for varying
values of shell thickness $\varepsilon$. The figure shows that
at some shell thickness values (dotted blue line), the SAWLC J-factor is
larger by about half an order of magnitude as compared to the WLC
J-factor (solid aqua line) for loop lengths $L\sim100$~bp. In Fig.~\ref{fig:WLC-removed-volume}(c),
we compare the J-factor for several SAWLCs to
their WLC counterparts, keeping $\varepsilon$ constant. The figure
shows that as the looping criterion for bond-center to bond-center
separation ($d_{min}+\varepsilon$) increases, the deviation from
the WLC J-factor prediction increases as well (compare red and red-dashed
to blue and blue-dashed lines). This result indicates that the deviation of the SAWLC J-factor from the WLC J-factor can be made to approach an order
of magnitude or more if the bond-center to bond-center
looping criterion is increased sufficiently.

\section{Discussion}

We carried out a simulation study of the looping or cyclization
J-factor for a self-avoiding worm-like chain (SAWLC) model, taking excluded volume effects into consideration for the first time. The distribution of DNA configurations generated
by our numerical algorithm was carefully tested by comparing to scaling
theory \cite{gennes_scaling_1979} and renormalization group \cite{chen_renormalization-group_1992}
results for the entropic regime, and indeed faithfully represented
the swollen chain partition function. In particular, our algorithm
was able to extract the universal exponent $\gamma$~\cite{gennes_scaling_1979},
and to reproduce the predicted $N^{-1.92}$ \cite{gennes_scaling_1979,sinclair_cyclization_1985}
J-factor power law for the entropic regime (i.e., where $L\gg l_{p}$). This
stronger power-law drop-off is a direct consequence of the self-avoiding assumption, under which the end-to-end separation grows like $\approx N^{0.5876}$
as compared with the ideal chain scaling of $\approx N^{0.5}$.
We further showed that the rate of convergence of the SAWLC J-factor power-law exponent to the asymptotic value of $-1.92$ is highly dependent on the
semi-flexible chain aspect ratio. This result implies that a highly
precise experimental determination of dsDNA aspect ratio can be made
by simply measuring this power-law exponent for various looping lengths.
Our model and its numerical implementation thus enabled the study of the swollen-chain J-factor in the elastic regime, where swollen-chain analysis
to our knowledge has not been carried out previously.

While the statistics in the elastic regime are dominated by elastic energy
considerations, our results indicate that the SAWLC model generates a small but detectible effect on the propensity
of the chain to loop. Our computations show that in this regime, the angle-independent probability density
function of the end-to-end vector $\mathbf{C}\left(\mathbf{r}\right)$
is almost identical for the WLC and SAWLC at points which are
accessible to both types of chains. However, there
is a range of end-to-end vector $\mathbf{r}$ which is inaccessible to the SAWLC (Fig.~\ref{fig:WLC-removed-volume}(a)). 
% This excluded volume results in a mild increase of the SAWLC J-factor as compared with the WLC J-factor. The increase is due to the anisotropic nature of the probability density function. Namely, the excluded volume for the SAWLC model overlaps the region for which the probability density is particularly low (The region enclosed in the light-green outline in Fig.~\ref{fig:WLC-removed-volume}(a), in the positive $\hat{z}$ direction). Equation ~\ref{eq:J_factor} suggests that the J factor is proportional to the average end-to-end probability density. The probability density averaged over the accessible volume increases, resulting in a larger propensity for looping.
The excluded volume of the SAWLC model leads to a redistribution of $\mathbf{C}_{SAWLC}(\mathbf{r})$, as shown in Fig.~\ref{fig:WLC-removed-volume}, with the combined weight of the $\mathbf{C}_{WLC}(\mathbf{r})$ within the excluded volume being shifted outside of the SAWLC excluded volume. This could result in an increase in $\int\limits _{\delta\mathbf{r}\,\mathrm{accessible}}\!\!\!\!\!\!\!\!\!\!\!\!\mathbf{C}_{SAWLC}(\mathbf{r})\mbox{d}\mathbf{r}$. Comparison of our numerical calculations of $\mathbf{C}(\mathbf{r})$ for both models (Fig.~\ref{fig:WLC-removed-volume}) shows that this redistribution of weight is roughly uniform over the entire range of $\mathbf{r}$. This uniformity is not surprising, since we could have generated the ensemble of short chains using importance sampling (IS, see Appendix ~\ref{sub:SUP-Importance-sampling}) and then discarded all self-crossing chains, thereby increasing the weight of all allowed chains uniformly. Moreover, if we assume that $\int\limits _{\delta\mathbf{r}\,\mathrm{inaccessible}}\!\!\!\!\!\!\!\!\!\!\!\!\!\!\mathbf{C}_{WLC}(\mathbf{r})\mbox{d}\mathbf{r}$ is redistributed uniformly over the $\mathbf{r}$ range, we obtain a relative increase in the J-factor on the order of $\Delta J/J\approx10^{-5}$. This is clearly not the main increase in the SAWLC J-factor observed in Fig.~\ref{fig:WLC-removed-volume}(b). Instead, the increase that we observe is likely due to the anisotropic nature of $\mathbf{C}(\mathbf{r})$. Namely, the excluded volume for the SAWLC model overlaps the region for which the probability density $\mathbf{C}_{WLC}(\mathbf{r})$ is particularly low (the region enclosed in the light-green outline in Fig.~\ref{fig:WLC-removed-volume}(a), in the positive $\hat{z}$ direction). Thus $\mathbf{C}_{WLC}(\mathbf{r})$ averaged over the part of $\delta\mathbf{r}$ accessible to the SAWLC model increases, which in turn leads to an increase in the J-factor (see Eq.~\ref{eq:J_factor}).
Interestingly, the particular choice of the boundary conditions for looping (e.g.
minimal bond-to-bond separation) determines the extent of the increase.
Consequently, the semi-flexible chain SAWLC J-factor generates the following picture as compared with the WLC J-factor: in the elastic regime, there is a slightly increased probability for
looping, while in the entropic regime this trend switches and the
SAWLC J-factor falls off with a power law greater than the $-1.5$ value predicted for the WLC.

Previous analysis have shown that the value of the J-factor at the
elastic regime can be made very sensitive to boundary conditions
and local deformations on the DNA~\cite{yan_statistics_2005,wiggins_exact_2005,vafabakhsh_extreme_2012}.
Our result adds another possible contribution that, together with the
previous explanations, could generate a cumulative effect that might explain
the seemingly anomalous bendability effect observed in cyclization
experiments~\cite{cloutier_dna_2005,wiggins_high_2006,vafabakhsh_extreme_2012}.

Using this insight, what are the experimental implications for looping
or cyclization experiments? First, experiments that aim
to test looping in the elastic regime will be strongly
dependent on boundary conditions. This implies that any observed deviation
from a consensus model should first be considered in the context of
better models that
simulate the effects of the particular boundary conditions.
Second, cyclization experiments that are carried out at larger polymer
lengths should be considered as the effects of the boundary conditions
become negligible for longer polymer contour lengths. In this regime,
cyclization measurements can yield accurate estimates for the persistence
length of the polymer and effective width or thickness based on measuring
the looping probability and comparing to the theoretical predictions
shown in Fig.~\ref{fig:Jfactor-power-law}. Finally, our model represents a realistic depiction of DNA molecules at length-scales that are relevant to biological regulation. We believe that it can form the basis for a more elaborate model that incorporates protein-DNA interactions.

\begin{acknowledgments}This research was supported by the Israel
Science Foundation through Grant No. 1677/12 and the Marie Curie Re-integration
Grant No. PCIG11-GA-2012-321675. YP would also like to thank support
provided by the Russel Berrie Nanotechnology Institute, Technion. SG thanks the Aly Kaufman Fellowship Trust for its support.
\end{acknowledgments}

\appendix

\section{\label{sub:SUP-Faithful-sampling}Faithful sampling}

\subsection{\label{sub:SUP-Importance-sampling}Importance sampling}

For the simple (non-self-avoiding) 3D freely-jointed chain (FJC) model,
\cite{phillips_physical_2009} faithful ensembles of polymer
chains can be generated computationally by random step-by-step sampling
of the angles of each successive link. In this method, a complete
$N$-mer chain is constructed in sequence, where the \textit{i}th
step samples the $\theta_{i},\phi_{i}$ angles to construct an \textit{i}-mer
chain. After sampling a sufficient number of chains, the partition
function can be evaluated from the $N_{c}$ sampled chains as follows:
\begin{equation}
Z_{N}=\sum_{j=1}^{N_{c}}e^{-\beta E_{j}}=N_{c},
\end{equation}
where $E_{j}=0$ in the FJC case. Any physical property $\left\langle f\right\rangle $
can be likewise computed:
\begin{equation}
\left\langle f\right\rangle =\frac{\sum\limits _{j=1}^{N_{c}}f\left(\left\{ \theta_{N},\phi_{N}\right\} _{j}\right)e^{-\beta E_{j}}}{\sum\limits _{j=1}^{N_{c}}e^{-\beta E_{j}}}=\frac{\sum\limits _{j=1}^{N_{c}}f\left(\left\{ \theta_{N},\phi_{N}\right\} _{j}\right)}{N_{c}} .
\end{equation}

When trying to collect an ensemble of plausible polymer configurations
for the WLC model, the simple sampling algorithm described above is
insufficient. Unlike the FJC model, in the WLC model link orientations
depend strongly on the elastic energy of bending (see Eq.~\ref{eq:WLC_Bending_E}).
Namely, sharp bends are highly improbable as compared to small perturbations
away from the non-bending minima. As a result, a uniform sampling
procedure of the relative link orientations will yield a disproportionate
number of improbable chains that are characterized by very large bending
energies, which will increase by orders of magnitude the computation
time necessary to obtain a statistically significant and representative
ensemble.

In order to alleviate this problem, and generate a faithful sampling
algorithm for the WLC model (i.e. $E^{hw}=0$), we employ a sampling
method termed importance sampling (IS) \cite{grassberger_pruned-enriched_1997}.
In IS we assign the following Boltzmann weight to the $i$th link:
\begin{equation}
\exp\left[-\beta E^{el}\left(\theta_{i},\phi_{i}\right)\right].
\end{equation}
We then sample the chain links according to the following probability
distribution:
\begin{equation}
p^{el}\left(\theta_{i},\phi_{i}\right)=\frac{\exp\left[-\beta E^{el}\left(\theta_{i},\phi_{i}\right)\right]}{\int\limits _{-1}^{1}\mbox{d}\cos\theta_{i}\int\limits _{0}^{2\pi}\mbox{d}\phi_{i}\exp\left[-\beta E^{el}\left(\theta_{i},\phi_{i}\right)\right]}.\label{eq:SUP-MC_importance_Sampling_p_i}
\end{equation}
As a result, the orientation angles that are more likely to occur
due to low bending energies will be chosen more frequently, thus reflecting
faithfully the underlying distribution. Furthermore, since each sampled
link appears in a chain of $N$ sampled links, the cumulative probability
to obtain each chain is a product of the individual link probabilities
sampled at every stage of the chain construction, as follows:
\begin{widetext}
\begin{eqnarray}
P_{tot}^{el}\left(\left\{ \theta_{N},\phi_{N}\right\} \right) & \equiv & \prod_{i=2}^{N}p^{el}\left(\theta_{i},\phi_{i}\right)=\prod_{i=2}^{N}\frac{\exp\left[-\beta E^{el}\left(\theta_{i},\phi_{i}\right)\right]}{\int\limits _{-1}^{1}\mbox{d}\cos\theta_{i}\int\limits _{0}^{2\pi}\mbox{d}\phi_{i}\exp\left[-\beta E^{el}\left(\theta_{i},\phi_{i}\right)\right]}\\
 & = & \frac{\exp\left[-\beta\sum\limits _{i=2}^{N}E^{el}\left(\theta_{i},\phi_{i}\right)\right]}{\int\limits _{-1}^{1}\mbox{d}\cos\theta_{1}\int\limits _{0}^{2\pi}\mbox{d}\phi_{1}\cdots\int\limits _{-1}^{1}\mbox{d}\cos\theta_{N}\int\limits _{0}^{2\pi}\mbox{d}\phi_{N}\exp\left[-\underset{i}{\sum}\beta E^{el}\left(\theta_{i},\phi_{i}\right)\right]}=\frac{\exp\left[-\beta E_{chain}\right]}{\int\limits _{all~configurations}\exp\left[-\beta E_{chain}\right]}.\nonumber
\end{eqnarray}
\end{widetext}
Thus, this method generates complete WLC chains according
to their statistical probabilities in the ensemble.\\
According to statistics theory,\cite{fristedt_modern_1996} if $X\,:\,\Omega\rightarrow\mathbb{R}$
is a random variable in some probability space $\left(\Omega,\mathcal{F},P\right)$
and one wishes to estimate the expected value of $X$ under $P$,
provided that one has random samples $x_{1},...,\, x_{n}$ generated
according to $P$, then an empirical estimate of $\mbox{E}[X;P]$
is
\begin{equation}
\hat{\mbox{E}}_{n}[X;P]=\frac{1}{n}\sum_{i=1}^{n}x_{i}.
\end{equation}
This simplifies the computation of any physical observable$\left\langle f\right\rangle $
for the IS to:
\begin{equation}
\left\langle f\right\rangle =\frac{1}{N_{c}}\sum\limits _{j=1}^{N_{c}}f\left(\left\{ \theta_{N},\phi_{N}\right\} _{j}\right).\label{eq:SUP-importance_sampling_<f>}
\end{equation}

\subsection{\label{sub:SUP-weighted-biased-sampling}Weighted-biased sampling}

The IS method can be used to generate chains for the SAWLC model by
following (\ref{eq:SUP-MC_importance_Sampling_p_i}) and then discarding
chains with overlapping links. However, this method is extremely inefficient
and leads to extreme sampling attrition \cite{sadanobu_continuous_1997}.

Naively one could replace Eq.~(\ref{eq:SUP-MC_importance_Sampling_p_i})
with
\begin{multline}
p_{i}\left(\left\{ \theta_{i},\phi_{i}\right\} \right)=\\
\frac{\exp\left[-\beta E^{el}\left(\theta_{i},\phi_{i}\right)\right]\Theta_{i}^{hw}\left(\left\{ \theta_{i},\phi_{i}\right\} \right)}{\int\limits _{-1}^{1}\mbox{d}\cos\theta_{i}\int\limits _{0}^{2\pi}\mbox{d}\phi_{i}\exp\left[-\beta E^{el}\left(\theta_{i},\phi_{i}\right)\right]\Theta_{i}^{hw}\left(\left\{ \theta_{i},\phi_{i}\right\} \right)},\label{eq:SUP-MC_Sampling_PDF}
\end{multline}
where $\Theta_{i}^{hw}\left(\left\{ \theta_{i},\phi_{i}\right\} \right)$
was defined previously in Eq.~(\ref{eq:Theta-HW}), and follow the
same (IS) procedure. However, since the hard-wall potential eliminates
a certain percentage of the possible configurations, a misrepresentation
of the sampled chains in the ensemble results, as discussed in \cite{rosenbluth_monte_1955}.
Therefore, in order to produce a faithful representation of the partition
function, each generated chain has to be assigned with a weight for
the excluded volume case.

In order to gain insight into the calculation of these weights and
their necessity, we consider the general case of self-avoiding chains. We denote
the probability of a chain $j$ with $N$ links to appear in a statistical
ensemble as $P_{j}^{S}$:
\begin{equation}
P_{j}^{S}=\frac{e^{-\beta E_{j}}}{Z_{N}^{S}},\label{eq:SUP-WBS-Z-stat}
\end{equation}
where $Z_{N}^{S}$ is the partition function of the statistical ensemble.
We denote the probability of a chain $j$ with $N$ links to be generated
with IS as $P_{j}^{IS}$. Due to the way the chain links are sampled:
\begin{widetext}
\begin{eqnarray}
P_{j}^{IS}=\prod_{i=2}^{N}p_{i}\left(\left\{ \theta_{i},\phi_{i}\right\} _{j}\right) & = & \prod_{i=2}^{N}\left(\frac{\exp\left[-\beta E^{el}\left(\theta_{i},\phi_{i}\right)\right]\Theta_{i}^{hw}\left(\left\{ \theta_{i},\phi_{i}\right\} _{j}\right)}{\int\limits _{-1}^{1}\mbox{d}\cos\theta_{i}\int\limits _{0}^{2\pi}\mbox{d}\phi_{i}\exp\left[-\beta E^{el}\left(\theta_{i},\phi_{i}\right)\right]\Theta_{i}^{hw}\left(\left\{ \theta_{i},\phi_{i}\right\} _{j}\right)}\right)\nonumber \\
 & = & \frac{e^{-\beta E_{j}}}{\prod_{i=2}^{N}\left(\int\limits _{-1}^{1}\mbox{d}\cos\theta_{i}\int\limits _{0}^{2\pi}\mbox{d}\phi_{i}\exp\left[-\beta E^{el}\left(\theta_{i},\phi_{i}\right)\right]\Theta_{i}^{hw}\left(\left\{ \theta_{i},\phi_{i}\right\} _{j}\right)\right)}\,.\label{eq:SUP-WBS-Z-IS}
\end{eqnarray}
\end{widetext} While the numerators in (\ref{eq:SUP-WBS-Z-IS}) and
(\ref{eq:SUP-WBS-Z-stat}) agree, the denominators differ. Furthermore,
the denominator in (\ref{eq:SUP-WBS-Z-IS}) is different for each
chain configuration, unlike the denominator in (\ref{eq:SUP-WBS-Z-stat}).
Thus, to remove the bias for chain $j$, it is sufficient to take
the denominator of (\ref{eq:SUP-WBS-Z-IS}) as the counter-weight
$W\left(\left\{ \theta_{N},\phi_{N}\right\} \right)$:
\begin{multline}
W\left(\left\{ \theta_{N},\phi_{N}\right\} \right)=\\
\prod_{i=2}^{N}\int\limits _{-1}^{1}\!\!\mbox{d}\cos\theta_{i}\int\limits _{0}^{2\pi}\!\!\mbox{d}\phi_{i}\exp\left[-\beta E^{el}\left(\theta_{i},\phi_{i}\right)\right]\Theta_{i}^{hw}\left(\left\{ \theta_{i},\phi_{i}\right\} \right),
\end{multline}
which allows to define the link-by-link counter-weight as
\begin{multline}
\omega_{i}\left(\left\{ \theta_{i},\phi_{i}\right\} \right)=\\
\int\limits _{-1}^{1}\mbox{d}\cos\theta_{i}\int\limits _{0}^{2\pi}\mbox{d}\phi_{i}\exp\left[-\beta E^{el}\left(\theta_{i},\phi_{i}\right)\right]\Theta_{i}^{hw}\left(\left\{ \theta_{i},\phi_{i}\right\} \right)=\\
\int\limits _{non-colliding\, directions}\mbox{d}\theta_{i}\mbox{d}\phi_{i}\exp\left[-\beta E^{el}\left(\theta_{i},\phi_{i}\right)\right],
\end{multline}
and the Rosenbluth factor as
\begin{equation}
W\left(\left\{ \theta_{N},\phi_{N}\right\} \right)=\prod_{i=1}^{N}\omega_{i}\left(\left\{ \theta_{i},\phi_{i}\right\} \right).\label{eq:SUP-MC_weight_fac-1}
\end{equation}

The partition function takes on the following form in the WBS case:
\begin{equation}
Z_{N}=\sum_{j=1}^{N_{c}}W\left(\left\{ \theta_{N},\phi_{N}\right\} _{j}\right),
\end{equation}
and the average value $\left\langle f\right\rangle $ of a physical
property is calculated as:
\begin{equation}
\left\langle f\right\rangle =\frac{\sum\limits _{j=1}^{N_{c}}f\left(\left\{ \theta_{N},\phi_{N}\right\} _{j}\right)W\left(\left\{ \theta_{N},\phi_{N}\right\} _{j}\right)}{\sum\limits _{j=1}^{N_{c}}W\left(\left\{ \theta_{N},\phi_{N}\right\} _{j}\right)}.\label{eq:SUP-MC_<f>}
\end{equation}

\section{Additional computations}

\subsection{Computation of power-law exponents\label{sub:SUP-power-law-exponent}}

To compute the power-law exponents $\nu$ of $\sqrt{\left\langle R^{2}\right\rangle }$ and $\chi$ of the J-factor in Sections~\ref{sub:Results-DNA-end-to-end-distance}
and \ref{sub:J-entropic}, respectively, we started off by taking a
logarithm of both the data and $L/b$. The power-law exponent could
then theoretically be read from the slope of the logarithm of the data
as a function of $\mbox{log}\left(L/b\right)$. The data, however,
are very sensitive to noise in the sampled data. Thus, we employed
a smoothing algorithm to deduce the slope of the data at each point:
We used a sliding window along the values of $L/b$. The range
of the plot covered by the sliding window was fitted to a linear function,
and the slope of the linear function was taken as the slope of the
plot at the center of the sliding window. The size of the sliding
window was increased as $L/b$ increased in order to account
for increased noise at greater values of $L/b$.

\subsection{A scaling theory derivation for the asymptotic power law for looping in the swollen coil regime\label{sub:SUP-asymptotic-J-power-law-derivation}}

The WLC model predicts that the J-factor will scale like
$\mathit{N^{-\frac{3}{2}}}$ in the entropic regime, when the length
of the polymer is much longer than the persistence length ($\mathit{N\gg l_{p}}$)
\cite{phillips_physical_2009}. Likewise, the swollen chain also exhibits
some power-law dependency in the entropic regime. The asymptotic behavior
of the SAWLC model in the entropic regime can be inferred from the
existing theory for a freely-jointed chain on a 3D lattice. In the
following derivation we rely on the arguments given in \cite{des_cloizeaux_lagrangian_1974,gennes_scaling_1979}.
We first assume that in the entropic regime, the probability of looping
scales as:
\begin{equation}
p_{loop}\left(N\right)\propto N^{-\alpha}.
\end{equation}
Since the probability for looping is defined as ratio of the number
of looped polymer configurations to the total number of polymer configurations,
we begin with the asymptotic expression for the total number of
available configurations for a self-avoiding chain of $N$ links \cite{mckenzie_polymers_1976}:
\begin{equation}
Z=\widetilde{z}^{N}N^{\gamma-1},
\end{equation}
where $\widetilde{z}$ is a number corresponding to the effective
lattice space that the chain is defined on (i.e. $\widetilde{z}=6$
for an ideal chain defined on a cubic lattice). Next, the mean square
end-to-end distance for a swollen chain follows a scaling law, originally
derived by Flory \cite{flory_statistical_1969}:
\begin{equation}
R_{f}\propto N^{\nu}.
\end{equation}
Combining these two results, it is possible to derive an expression for the total number of looped states for
a self-avoiding random-walk chain. This result is derived for a grid
with pixels of size $a$, such that a loop is formed when the two
ends are separated by a distance $a$. Thus, on a grid, a self-avoiding
random-walk chain will form a closed polygon of $N+1$ edges, which
implies \cite{gennes_scaling_1979}
\begin{equation}
Z_{loop}\propto\widetilde{z}^{N}\left(\frac{a}{R_{f}}\right)^{3},
\end{equation}
with a normalizing term which takes into account that the loop can
be closed anywhere within a sphere of radius which is roughly equal
to the Flory radius . Given this expression, we can now write the
expression for the probability for looping in the asymptotic regime \cite{gennes_scaling_1979}:
\begin{equation}
p_{loop}\left(N\right)\equiv\frac{1}{a^{3}}\frac{Z_{loop}}{Z}\propto\frac{1}{R_{f}^{3}\, N^{\gamma-1}}\propto N^{-\left(3\nu+\gamma-1\right)},
\end{equation}
which is the desired power law. Finally, using the RG derived values
for $\nu=0.5876$ and $\gamma=1.1568\simeq7/6,$ \cite{caracciolo_high-precision_1998,clisby_self-avoiding_2007,clisby_accurate_2010} we obtain
\begin{equation}
\chi=3\nu+\gamma-1=1.9196\,.
\end{equation}
The universal nature of exponent $\gamma$ implies that the J-factor
power law for semi-flexible swollen chains ($\frac{w}{b}<1$) can
be predicted simply from the end-to-end exponent $\nu$.

\bibliographystyle{unsrt}
\bibliography{cyc_bibliography}

\end{document}